\documentclass[useAMS,usenatbib]{mn2e}
\usepackage{times}

\usepackage{amssymb}
\usepackage{psfig}

\title[Planetary nebulae as stellar population tracers]
{Planetary nebulae as tracers of galaxy stellar populations}
\author[A. Buzzoni, M. Arnaboldi \& R.L.M. Corradi]{Alberto Buzzoni$^1$, Magda Arnaboldi$^{2,3}$, \&
Romano L.M. Corradi$^{4,5}$\\
$^1$ INAF - Osservatorio Astronomico di Bologna, Via Ranzani 1, 40127 Bologna, Italy; 
\,{\sf e-mail: buzzoni@bo.astro.it}\\
$^2$ ESO - Karl-Schwarzschild-Str. 2, 85748 Garching bei M\"unchen, Germany;\,{\sf e-mail: marnabol@eso.org}\\
$^3$ INAF - Osservatorio Astronomico di Torino, Via Osservatorio 20, 10025 Pino Torinese (To), Italy\\
$^4$ ING - Isaac Newton Group of Telescopes, A.P. 321, 38700 Santa Cruz de La Palma, Spain;\,{\sf e-mail: rcorradi@ing.iac.es}\\
$^5$ IAC - Instituto de Astrof\'\i sica de Canarias, Via L\'actea s/n, 38200, La Laguna, Tenerife, Spain
}

\begin{document}

\date{Accepted ... Received ... in original form}

\pagerange{\pageref{firstpage}--\pageref{lastpage}} \pubyear{2006}

\maketitle

\label{firstpage}

\begin{abstract}

We address the general problem of the luminosity-specific planetary
nebula (PN) number, better known as the ``$\alpha$'' ratio, given by
$\alpha = {N_{\rm PN}/L_{\rm gal}}$, and its relationship with age and
metallicity of the parent stellar population.  Our analysis relies on
population synthesis models, that account for simple stellar
populations (SSPs), and more elaborated galaxy models covering the
full star-formation range of the different Hubble morphological types.
This theoretical framework is compared with the updated census of the
PN population in Local Group galaxies and external ellipticals in the
Leo group, and the Virgo and Fornax clusters.

The main conclusions of our study can be summarized as follows:

{\it i)} according to the Post-AGB stellar core mass, PN lifetime in a SSP
is constrained by three relevant regimes, driven by the nuclear
($M_{\rm core} \gtrsim 0.57~M_\odot$), dynamical ($0.57~M_\odot
\gtrsim M_{\rm core} \gtrsim 0.55~M_\odot$) and transition
($0.55~M_\odot \gtrsim M_{\rm core} \gtrsim 0.52~M_\odot$) timescales.
The lower limit for $M_{\rm core}$ also sets the minimum mass for 
stars to reach the AGB thermal-pulsing phase and experience the PN
event;

{\it ii)} mass loss is the crucial mechanism to constrain the value of
$\alpha$, through the definition of the initial-to-final mass relation
(IFMR). The Reimers mass-loss parameterization, calibrated on
Pop~II stars of Galactic globular clusters, poorly reproduces the
observed value of $\alpha$ in late-type galaxies, while a better fit
is obtained using the empirical IFMR derived from white-dwarf
observations in the Galaxy open clusters;

{\it iii)} the inferred PN lifetime for Local Group spirals and
irregulars exceeds 10\,000~yr, which suggests that $M_{\rm core}
\lesssim 0.65~M_\odot$ cores dominate, throughout;

{\it iv)} the relative PN deficiency in elliptical galaxies, and the observed
trend of $\alpha$ with galaxy optical colors support the presence of a
prevailing fraction of low-mass cores ($M_{\rm core} \lesssim
0.55~M_\odot$) in the PN distribution, and a reduced visibility
timescale for the nebulae as a consequence of the increased AGB
transition time.  The stellar component with $M_{\rm core} \lesssim
0.52~M_\odot$, which overrides the PN phase, could provide 
an enhanced contribution to hotter HB and Post-HB
evolution, as directly observed in M~32 and the bulge of M~31.  This
implies that the most UV-enhanced ellipticals should also display the
lowest values of $\alpha$, as confirmed by the Virgo cluster
early-type galaxy population;

{\it v)} any blue-straggler population, invoked as
progenitor of the $M_{\rm core} \gtrsim 0.7~M_\odot$ PNe in
order to preserve the constancy of the bright luminosity-function
cut-off magnitude in ellipticals, must be confined to a 
small fraction (few percents at most) of the whole galaxy PN
population.

\end{abstract}


\section{Introduction}\label{intro}

Diffuse intracluster luminosity \citep{uson} and other faint surface-brightness 
features detected at large distances from the center of isolated and cluster galaxies
\citep{hui,mihos} may provide a valuable 
record of the mechanisms that led to the assembly and formation of the
cosmic structures at the different hierarchical scales.

Stellar streams, like the case of Sgr and CMa in the Milky Way
\citep[e.g.][]{martin}, and ``runaway'' halo stars
\citep{blaauw,keenan,allen} are, in this sense, excellent examples of
the effect of galaxy tidal interactions and dynamical relaxation
processes, that spread stars well outside the bright body of their
parent systems.  Furthermore, the evidence of quiescent on-going star
formation in low-density environments, such as dwarf or very low
surface-brightness galaxies \citep{hunter,caldwell,cellone}, indicates
that the transition between coherent stellar systems and diffuse
background might be not so sharp.

In this framework, the study of Planetary Nebulae (PNe) is of special
interest as these objects are efficient tracers of their underlying
parent stellar population even in those regions where the stellar plot
is too faint to be detected against the night-sky brightness
\citep{arnaboldi}. Relying on standard narrow-band imagery, PNe have
been confidently detected out to 70~kpc from the center of Cen~A
\citep{hui} and other bright galaxies in the Virgo and Fornax
clusters, about 15~Mpc away
\citep[][]{arnaboldi,arnaboldi04b,feld03,feld04,peng,aguerri}. On the
other hand, optimized multi-slit imaging recently pushed this distance
limit even farther, reaching the Coma cluster PNe, at about
100~Mpc distance \citep[][]{gerhard}.

One open question in this approach is the link between the size of the
PN component and the sampled (bolometric) luminosity of its parent
stellar population.  This ratio, often referred to in the literature as the
``luminosity-specific PN number density'' or, shortly, the ``$\alpha$ ratio''
\citep{jacoby}, provides the amount of light associated to any
observed PN sample, and it is the first assumption when using PNe as
tracers of the spatial distribution and motions of the parent stars.
Observationally, there is strong evidence for $\alpha$ to
correlate with galaxy color, the reddest ellipticals
being a factor 5 to 7 poorer in PNe per unit galaxy luminosity than
spirals \citep{peimbert,hui}. This trend is at odds with a nearly
``universal'' PN luminosity function (PNLF), as observed for galaxies
along the whole Hubble morphological sequence \citep{ciardullo}.

Population synthesis models offer a useful reference tool, in this regard,
as they allow us to probe the theoretical relationship of $\alpha$ with the
different distinctive parameters of a stellar population, thus
overcoming most of the observational uncertainties related to
systematic bias, incomplete sampling etc.. In addition, a theoretical
assessment of the $\alpha$ ratio does not require, in principle, any
exact knowledge of the PNLF, still largely uncertain at its faint magnitudes
\citep{jdm,jacoby05}.

In this paper we want to study the luminosity-specific PN number
density from a more general approach based on a new set of stellar
population models.  Our calculations rely on the \citet[][hereafter
B89]{b89} and \citet{b95} synthesis code, including the recent
extension to template galaxy models of composite star formation
\citep[][]{b02,b05}.  Our predictions for the $\alpha$ ratio will also
be compared with the observed PN population of Local Group galaxies,
in order to validate our adopted scenario for Post-asymptotic giant
branch (PAGB) evolution. We will also address the puzzling problem of the poorer 
PN population in elliptical galaxies \citep{hui}: this very interesting feature might
be the signature of a phase transition related to horizontal branch
(HB) evolution, with important consequences on the expected
ultraviolet evolution of galaxies and other stellar systems in
metal-rich environments.

Our paper will be organized as follows: in Sec.~\ref{sec:s2} we shall
present the simple stellar population (SSP) theory, and discuss the
relevant parameters that constrain $\alpha$.  Section~\ref{galx} is
devoted to the expected theoretical values of the $\alpha$ parameter
for galaxies of different morphological type. In Sec.~\ref{obs} we
compare the theoretical results with observations of the PN population
in Local Group and external galaxies in the Leo group, the Virgo
and Fornax clusters.  In Sec.~\ref{ell}, we study the
correlation between $\alpha$ and other spectrophotometric indices for
early-type galaxy diagnostic, like the Lick Mg$_2$ index and the
$(1550-V)$ ultraviolet color. Finally, in Sec.~\ref{end} we
summarize the constraints obtained for the PAGB evolution and the
relevant timescale for PN evolution.

\section{Simple stellar population theory}
\label{sec:s2}

The SSP theory developed by \citet[][hereafter RB86]{rb86} and
\citeauthor*{b89} allows us to compute the value of $\alpha$ for a coeval
and chemically homogeneous stellar aggregate.  Assuming a standard
power-law IMF such as $N_*(M_*) \propto M_*^{-s}\,dM_*$, the expected
number of stars that populate the {\it ``j''}-th Post-main sequence
(PMS) phase, of duration $\tau_j$, can be written as:
\begin{equation}
N_j =  A M_{\rm TO}^{-s}\, |\dot{M}_{\rm TO}|\, \tau_j,
\label{eq:1}
\end{equation}
where $\dot{M}_{\rm TO}$ is the time
derivative of the MS turn off (TO) mass.  The scaling
factor $A$ in eq.~(\ref{eq:1}) accounts for the total mass of the SSP
(cf.\ \citeauthor*{b89}), and relates to the total luminosity of the
stellar population, so that $N_j \propto L_{\rm
tot}\,\tau_j$.  

When this approach is applied to the PN phase, it becomes:
\begin{equation}
N_{\rm PN} = {\cal B}\, L_{\rm tot}\, \tau_{\rm PN},
\label{eq:2}
\end{equation}
where ${\cal B} = A M_{\rm TO}^{-s} |\dot{M}_{\rm TO}|/L_{\rm tot}$ is
the so called ``specific evolutionary flux'' (see \citeauthor*{rb86}
and \citeauthor*{b89}), and $\tau_{\rm PN}$ is the PN {\it visibility}
lifetime, i.e.\ the time for the nebula to be detectable in [O{\sc
iii}] and/or H$\alpha$ surveys. The value of the ``luminosity-specific
PN number'' $\alpha$ simply derives from eq.~(\ref{eq:2}) as
\begin{equation}
\alpha =  {N_{\rm PN} \over L_{\rm tot}} = {\cal B}\,\tau_{\rm PN}.
\label{eq:3}
\end{equation}
Therefore two parameters, that is ${\cal B}$ and $\tau_{\rm PN}$, set the value of 
the luminosity-specific PN number for a given SSP.

\subsection{The ``specific evolutionary flux'' ${\cal B}$}

This parameter links the PN rate to the evolutionary properties of the
parent stellar population. It contains, in fact, {\it i)} the physical
``clock'' of the population (as both $M_{\rm TO}$ and $\dot{M}_{\rm
TO}$ depend on time), and {\it ii)} the IMF dependence, because the
ratio $A/L_{\rm tot}$ closely scales with the SSP $M/L$ ratio, thus
responding to the mass distribution of the composing stars.

Following \citet{b98}, we can further arrange our definition of $\cal B$ and write
\begin{equation}
{\cal B} = {{A M_{\rm TO}^{-s} |\dot{M}_{\rm TO}|}\over {L_{\rm tot}}} = {{\cal L} \over {\rm PMS~fuel}}. 
\label{eq:4}
\end{equation}
The r.h. side of eq.~(\ref{eq:4}) derives from the \citeauthor*{rb86}
``Fuel consumption theorem'' and identifies the PMS
contribution to the total SSP luminosity (${\cal L} = L_{\rm
PMS}/L_{\rm tot}$), as well as the {\it absolute} nuclear fuel spent during
PMS evolution by stars with $M_* = M_{\rm TO}$.

\begin{figure}
\centerline{
\psfig{file=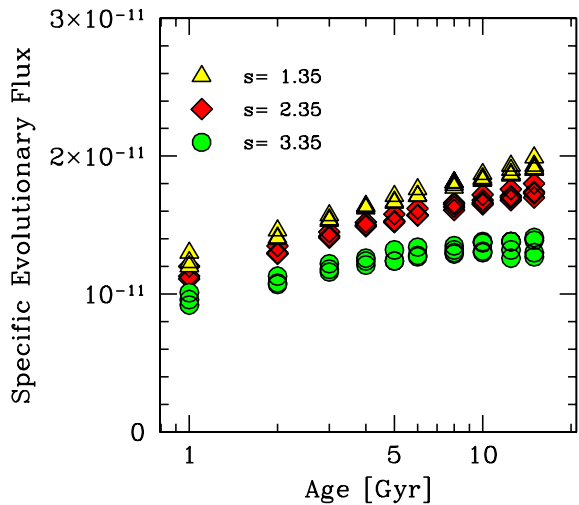,width=\hsize,clip=}
}
\caption{Specific evolutionary flux $\cal B$, from eq.~(\ref{eq:4}),
for \citeauthor*{b89} SSP models.  Different metallicity sets (between
$Z \sim 1/20$ and $2~Z_\odot$) are overplotted.  In addition to the
Salpeter case ($s = 2.35$), other IMF power-law coefficients are
explored, as labeled on the plot. The value of $\cal B$ is given in
units of $L_\odot^{-1}$\,yr$^{-1}$.  }
\label{bb}
\end{figure}

As pointed out by \citeauthor*{rb86}, $\cal B$ does not change much 
with metallicity or IMF slope; this is shown in Fig.~\ref{bb}, where
we display the specific evolutionary flux,  
according to \citeauthor*{b89} SSP models, for different IMF slopes and
metallicity. One can see that models with $Z = 1/20$ to 2~Z$_\odot$ and IMF
slopes from 1.35 to 3.35 changes $\cal B$ by only a factor of two, at most,
over a wide range of SSP age, from 1 to 15 Gyr.

The dependence of $\alpha$ on the IMF slope is through the change in
$\cal L$, i.e.\ the relative PMS luminosity contribution in
eq.~(\ref{eq:4}).  In general, a giant-dominated SSP will display a
larger value of $\alpha$ because giant stars become more important in
a ``flatter'' IMF and $\cal L$ tends to increase, compared to the
Salpeter $s=2.35$ case. On the other hand, a younger age acts in the
sense of decreasing the PN rate per unit SSP luminosity, because the
MS luminosity contribution increases, the absolute amount of ``fuel''
burnt by PMS stars increases (cf., e.g.\ Fig.~3 in
\citeauthor*{rb86}), and $\cal L$ decreases.

\subsection{The PN lifetime $\tau_{PN}$}
\label{fc}

This quantity depends both on the chemio-dynamical
properties of the ejected material, giving rise to the nebular
envelope, and on the stellar core-mass evolution, which is responsible
for the ``firing up'' of the nebular gas.

The first self-consistent theoretical model of PN core evolution is
due to \citet{paczynski}, then followed by contributions from
\citet{scho81,scho83}, \citet{gorny}, \citet{vw}, \citet{leti1},
\citet{leti2}, and more recently by \citet{marigo} and
\citet{perinotto04}.  Theory predicts that the PN properties
strongly depend on the core-mass distribution of PAGB remnants, the
latter being the result of the initial-to-final mass relation
(hereafter IFMR), as modulated by metallicity and mass loss.  The
so-called ``transition time'' for PAGB stars to reach the
high-temperature region ($T_{\rm eff} \gtrsim 30\,000$~K), after the
nebula ejection, may also play a role on the emission properties and
the timescale of the PN event.

\begin{figure}
\centerline{
\psfig{file=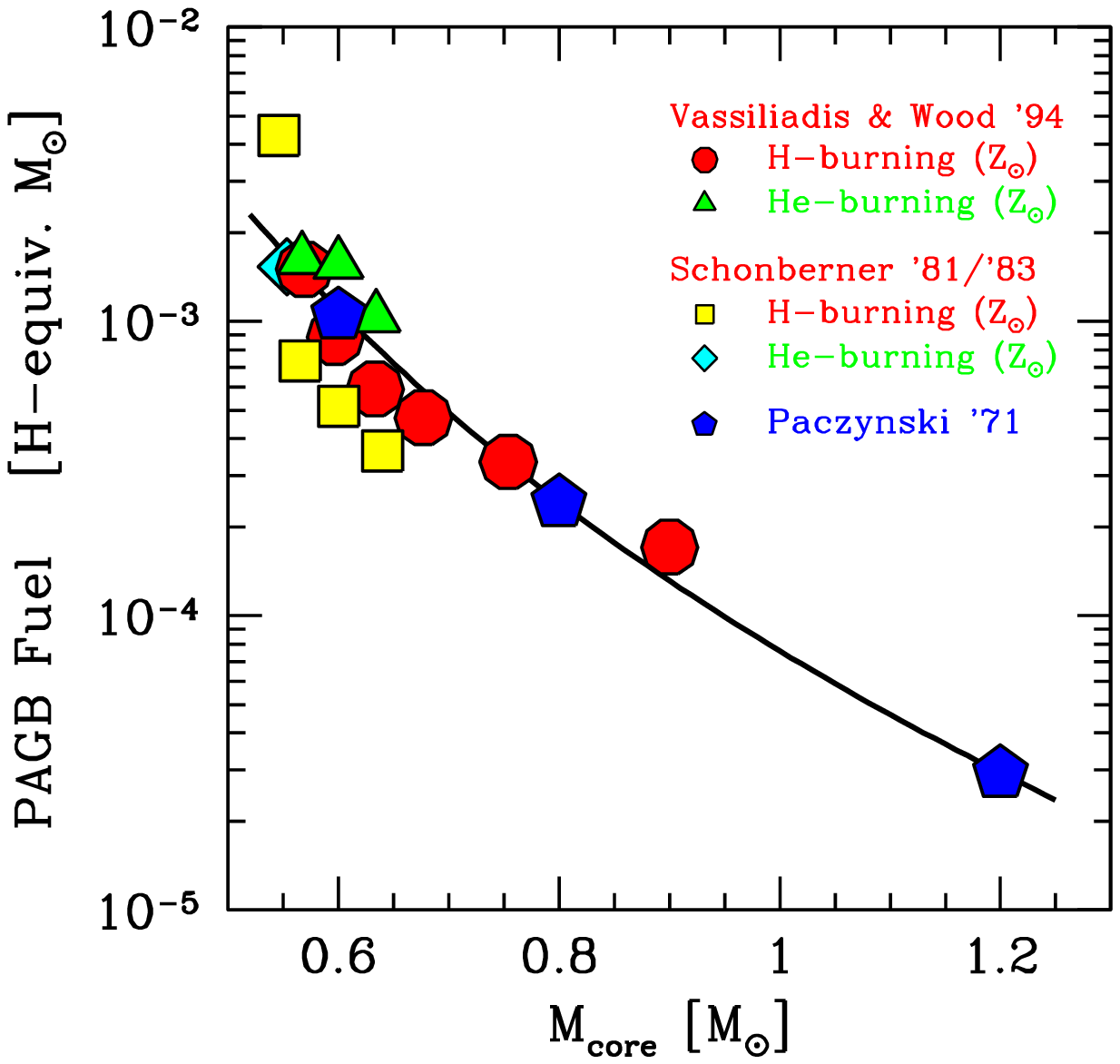,width=0.86\hsize,clip=}
}
\caption{Theoretical fuel consumption for stars along the PAGB evolution
according to different model sets: \citet[][pentagon
markers]{paczynski}, \citet[][squares and
rhombs]{scho81,scho83},\citet[][dots and triangles]{vw}.  The different
markers for the same model source refer to the prevailing case of a H
or He thermal pulse terminating the AGB evolution, as labeled. Fuel is
expressed in Hydrogen-equivalent solar mass, i.e. 1\,g of H-equivalent
mass~= $6~10^{18}$~ergs (cf.\ \citeauthor*{rb86}), and a solar
metallicity is assumed in the models.  A smooth analytical relation
matching the data, according to eq.~(\ref{eq:fuel}) is plotted as a solid curve.}
\label{fuel}
\end{figure}

A simple estimate of PN core lifetime can be derived from the
energetic budget available to PAGB stars. This is computed in
Fig.~\ref{fuel}, based on a model collection from \citet{paczynski},
\citet{scho81,scho83} and \citet{vw}.  Indeed, one sees that \citet*{paczynski}
original results consistently meet the more recent and sophisticated stellar
tracks, like those of \citet{vw}, that account for stellar envelope
ejection either in the case of a He or H thermal pulse at the tip of
AGB evolution.

\begin{figure*}
\begin{minipage}{0.69\hsize}
\psfig{file=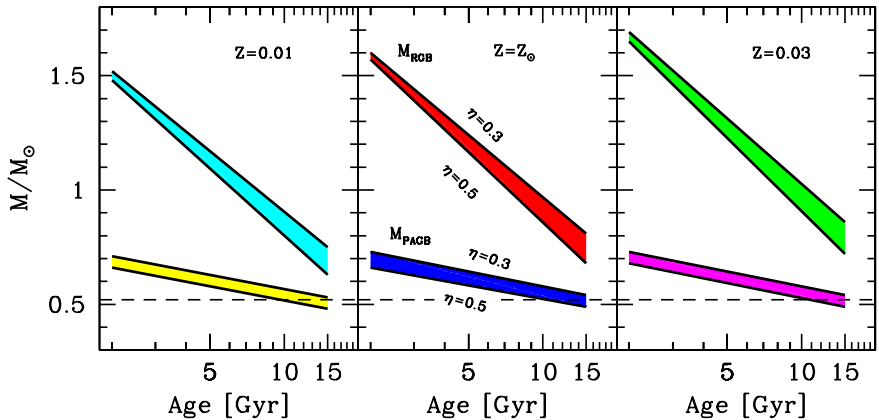,width=\hsize,clip=}
\end{minipage}
\hfill
\begin{minipage}{0.30\hsize}
\caption{Time evolution of the stellar mass at some tipping points
across the H-R diagram for SSPs of different metallicity $Z$, about
the solar value, as labeled in each panel.  Upper strips in each panel
are the theoretical loci for stellar mass at the tip of the RGB
evolution, ($M_{\rm RGB}$) according to a Reimers mass loss parameter
in the range $0.3 \le \eta \le 0.5$, as labeled in the middle panel,
for general reference. Lower strips mark the locus for stellar mass at
the onset of PAGB evolution ($M_{\rm PAGB}$), again for the same
reported range of the mass loss parameter $\eta$. The minimum mass for
stars to reach the AGB thermal pulsing phase (and eventually produce a
PN) is marked by the dashed line, according to \citet{dorman} and
\citet{blocker}.  }
\label{mpn}
\end{minipage}
\end{figure*}

From \citeauthor*{b89}, an analytical function that reproduces the
PAGB fuel consumed as a function of the core mass for the different
theoretical data sets of Fig.~\ref{fuel} is
\begin{equation}
{\rm PAGB~fuel} = (M_{\rm core}/0.163)^{-5.22} \qquad [H~M_\odot],
\label{eq:fuel}
\end{equation}
where $M_{\rm core}$ is in solar unit and fuel in Hydrogen-equivalent
solar mass (cf.\ Fig.~\ref{fuel} for details).  The PAGB stellar
lifetime can be defined as
\begin{equation}
\tau_{\rm PAGB} \simeq ({\rm PAGB~fuel} / \ell_{\rm PAGB}),
\label{eq:time}
\end{equation}
being $\ell_{\rm PAGB}$ the luminosity of the stellar core at the
onset of the nebula ejection \citep[see e.g.][]{paczynski}.  In
general, $\tau_{\rm PAGB}$ is shorter than the estimated dynamical
timescale for the nebula evaporation, which is about $\tau_{\rm dyn}
\simeq 30\,000$~yr \citep[cf.][]{scho83,phillips}, and this implies
that for SSPs of young and intermediate age $\tau_{\rm PN} \simeq
\tau_{\rm PAGB}$.

On the other hand, when the SSP age increases, this assumption may no
longer hold, and at some point, when PAGB stellar core mass decreases
and $M_{\rm core} \lesssim 0.57\,M_\odot$, $\tau_{\rm PAGB}$ starts to
exceed $\tau_{\rm dyn}$.  Henceforth, the PN evolution is driven by the
dynamical timescale, reduced by the transition time.


\subsection{PAGB core mass and PN evolution}\label{mcore}

As $M_{\rm core}$ constrains both the fuel and PAGB luminosity, from
which $\tau_{\rm PAGB}$ derives,\footnote{In force of
eq.~(\ref{eq:fuel}), and also considering that $\ell_{\rm PAGB}
\propto M_{\rm core}$ in eq.~(\ref{eq:time}) \citep{paczynski}, we
have that $\tau_{\rm PAGB} \propto M_{\rm
core}^{-6.22}$.\label{nota1}} it is clear that  mass loss
mechanisms, at work along the giant-branch evolution, set the leading
parameters for PN evolution. In this sense, it is of
paramount importance to establish a suitable relationship between the
initial (i.e.\ $M_{\rm i} \equiv M_{\rm TO}$) and final ($M_{\rm f}
\equiv M_{\rm PAGB}$) mass of stars along the whole SSP evolution.

A firm settlement of the IFMR is a long-standing problem, that has been addressed both
theoretically (\citealp{iben}, hereafter IR83, \citealp{dominguez}, \citealp{girardi}) and
observationally (\citealp{w83,w87,weidemann}, hereafter W00; \citealp{claver,kalirai}).

\subsubsection{Mass loss and $M_{\rm core}$: the theoretical approach}

A direct evaluation of the mass loss, according to the \citet{reimers}
theoretical parameterization, is shown in Fig.~\ref{mpn}.  In this
figure, we plot the expected value of stellar masses at some tipping
points of SSP evolution, for three relevant values of metallicity
around the solar value. In particular, $M_{\rm RGB}$ is the mass of
stars at the end of red giant branch (RGB) evolution, which occurs
after the first important mass-loss episode experienced by stars at
the low-gravity low-temperature regime; $M_{\rm PAGB}$ is then the
mass of the stars leaving the AGB, after the second stronger mass-loss
episode.  The trend of both quantities is tracked vs.\ SSP age, for a
range of the Reimers mass-loss parameter, $\eta$, between 0.3 and
0.5.\footnote{A value of $\eta \simeq 0.4 \pm 0.1$ is typically
required in order to reproduce the observed colour-magnitude (c-m) diagram of old
Galactic globular clusters \citep{fusipecci}.}

Along the SSP evolution shown in Fig.~\ref{mpn}, a substantial fraction (up
to 50\%) of the stellar mass is lost during the AGB phase (cf.\
the difference $\Delta M_* = M_{\rm RGB} - M_{\rm PAGB}$ on the plots); 
furthermore, with increasing age, the core mass of
PAGB stars approaches (or even crosses) the limit of $M_{\rm PAGB} \simeq
0.52$~M$_\odot$, which is the minimum mass required by models for stars
to experience the so-called AGB ``thermal-pulsing'' phase
\citep[][]{dorman,blocker}. During this phase, stars venture in the
region of Mira variables and end their AGB evolution with a quick
envelope ejection, likely driven by dynamical instability (the
so-called ``superwind phase'' of \citealp{renzini}), and the
subsequent PN stage (see \citealp{pac70}, and \citeauthor*{iben}, 
for an exhaustive discussion of the process and its
variants). We shall discuss the implication of an inhibited AGB phase
on PN evolution in what follows.

\subsubsection{Mass loss and $M_{\rm core}$: the empirical approach}\label{caseab}

The IFMR can be derived empirically from the mass estimate 
of observed white dwarfs in nearby open clusters (like Hyades or
Praesepe), coupled with the value of $M_{\rm TO}$ and cluster age,
as obtained from isochrone fit of the cluster c-m diagram. 

Quoting \citeauthor*{weidemann}, there is evidence that observed white dwarf 
masses, for low- and intermediate-mass stars, 
``coincide almost exactly with the new theoretical predictions of the
core masses at the beginning of the thermal pulsing AGB'' ($M_{\rm
TP}$). In addition, ``it is presumed and supported weakly by the
empirical data that this closeness of the final mass to the
first-thermal pulse core mass relation continues also to higher
initial masses'', possibly up to the limit of SN onset (roughly placed
about $7~M_\odot$).

\begin{figure}
\centerline{
\psfig{file=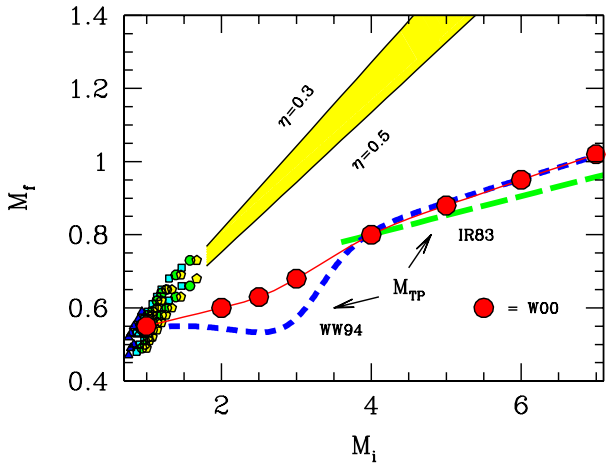,width=\hsize,clip=}
}
\caption{The initial-to-final mass relation according to different
calibrations.  The solid strip is the theoretical relation of
\citeauthor*{iben} for a standard mass loss parameter $\eta$ in the
range between 0.3 and 0.5, as labeled on the plot. Small dots report
the individual values as from the \citeauthor*{b89} SSP models of
Table~\ref{alpha} and the same Reimers parameters.  Short- and
long-dashed curves are the theoretical loci for stars to set on the
AGB thermal pulsing phase ($M_{\rm TP}$), according to \citeauthor*{iben} and
\citet{ww94} (WW94). Finally, big dots and solid curve report the
\citeauthor*{weidemann} empirical relation based on the mass estimate
of white dwarfs in Galactic open cluster. }
\label{mfin}
\end{figure}

The first claim of this analysis is confirmed by Fig.~\ref{mfin},
where we compare the \citeauthor*{weidemann} IFMR and the theoretical
estimates of $M_{\rm TP}$ from an updated set of stellar tracks for
Pop I stars by \citet{ww94} and from the original analytical relation
by \citeauthor*{iben} for intermediate-mass stars, namely
\begin{equation}
M_{\rm TP} = 0.59 +0.0526\,M_{\rm i}.
\label{eq:tp}
\end{equation}

It is also clear from Fig.~\ref{mfin} that, for a standard range of
Reimers mass-loss parameters fitted to Galactic globular clusters, the
\citeauthor*{iben} theoretical IFMR overestimates the value of $M_{\rm f}$
for young ($t \lesssim 2$~Gyr) SSPs, requiring a value of $\eta
\gg 1$ to match the \citeauthor*{weidemann} empirical relation.

Given the discrepancy between the predicted $M_{\rm core}$ based on the
standard mass loss theory {\it \`a la} \citet{reimers} and the
observations by \citeauthor*{weidemann}, we compute the $\tau_{\rm
PAGB}$ values according to {\it a)} a theoretical IFMR according to
\citeauthor*{b89} SSP models with $\eta = 0.3$, extended to younger
ages through the \citeauthor*{iben} relation
\begin{equation}
M_{core} = 0.53\,\eta^{-0.082} + 0.15\,\eta^{-0.35}(M_{\rm TO} -1),
\end{equation}
and {\it b)} the \citeauthor*{weidemann} empirical IFMR, 
as displayed in Fig.~\ref{mfin}.

\subsubsection{A critical core mass range: $0.52\,M_\odot \leq M_{\rm core} \leq 0.55\,M_\odot$}\label{casec}

When the core nuclear lifetime ceases to be the driving parameter for
PN visibility (that is for $M_{\rm core} \lesssim 0.57~M_\odot$), we
have that $\tau_{\rm PN} \simeq (\tau_{\rm dyn} - \tau_{\rm tt})$,
where $\tau_{\rm tt}$ is the transition time of the stellar core to
become hot enough such as to ``fire up'' the nebula. Models show that,
for $M_{\rm core} \lesssim 0.55~M_\odot$, $\tau_{\rm tt}$ abruptly
increases from its typical value range of 200-2000~yr up to exceeding
$\tau_{\rm dyn}$ \citep{scho83,vw}. In this case, by the time the stellar
core is there ready to heat, the gaseous shell has already evaporated,
thus preventing the nebula ignition.  The dynamical timescale itself
is slightly influenced by mass loss as a slower envelope expansion,
driven by radiation pressure, is expected if $\eta$ increases such as
to terminate AGB evolution at lower luminosity
\citep[][]{vw,marigo,villaver}.

{\it In general, this scenario leads to conclude that in old SSPs 
PN visibility might be greatly reduced or even fully inhibited as a consequence of 
a delayed hot-PAGB evolution of the stellar core} \citep{leti2}.

In addition, one has also to account for a further critical threshold
in the evolutionary framework when stars escape the AGB
thermal-pulsing phase, for $M_{\rm core} \lesssim 0.52~M_\odot$. The lack
of a full AGB development leads to a range of Post-HB evolutionary 
paths,\footnote{From the physical point of view, this would
correspond to the He+H double-shell burning regime for low- and
intermediate-mass stars.} as discussed in detail by
\citet{greggio}. One relevant case, in this regard, is that of {\it
``AGB-manqu\'e''} stars, that directly set on the
high-temperature white-dwarf cooling sequence after leaving the HB,
thus missing, partially or {\it in toto}, the AGB phase. First
important hints of this non-standard evolutionary framework can be
found in the original work of \citet{gingold}, and a number of later
contributions tried to better assess the critical parameters (mass
loss {\it in primis}) involved in the process
\citep[][]{castellani92,castellani,castellani06,dorman,dcruz,yi}.

The impact of this composite scenario on the PN formation mechanisms
is probably still to be fully understood; it seems clear, however,
that a full completion of AGB evolution, up to the thermal-pulsing
luminosity range, is the most viable step to culminate with the nebula
event \citep[see also][for a more elaborated discussion in this regard]{kwok}.

\subsection{The luminosity-specific PN number in Simple Stellar Populations}
\label{mc}

According to the previous discussion, the luminosity-specific PN number in a 
SSP can eventually be written as
\begin{equation}
\alpha = \left\{
\begin{array}[]{ll}
{\cal B}\,{\rm min} [ \tau_{\rm PAGB}, (\tau_{\rm dyn} - \tau_{tt}) ],
& \qquad {\rm if}~\tau_{\rm tt} \le \tau_{dyn} \\ 0 & \qquad {\rm
otherwise}.
\end{array}
\right.
\label{eq:alfa}
\end{equation}

In general, $\tau_{\rm dyn}$ {\it sets a safe upper limit to the $\alpha$ value}.
For a Salpeter IMF, and from the data of Fig.~\ref{bb}, we derive
\begin{equation}
\alpha_{\rm max} \simeq 1.8\,10^{-11} \times 30\,000 = {{1~{\rm PN}}\over {1.85\,10^6~{\rm L}_\odot}}.
\label{eq:amax}
\end{equation}
This value is virtually independent from metallicity.

\begin{figure}
\centerline{
\psfig{file=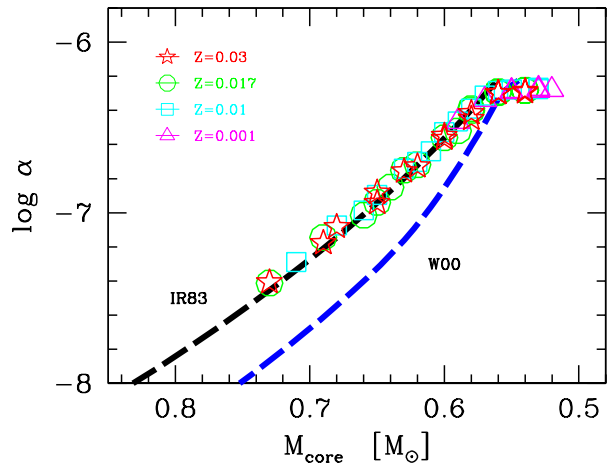,width=\hsize,clip=}
}
\caption{The luminosity-specific PN number for SSP models of
  Table~\ref{alpha} (both for $\eta = 0.3$ and 0.5 and different
  metallicity, as reported top left) compared to the PAGB stellar core
  mass. Overplotted are also the expected calibration assuming the
  theoretical IFMR of \citeauthor*{iben} and the empirical one from
  \citeauthor*{weidemann}, as labeled.  Note the clean relationship in
  place, with $M_{\rm core}$ being the leading parameter to constrain
  $\alpha$.  }
\label{mcr}
\end{figure}

\begin{figure}
\centerline{
\psfig{file=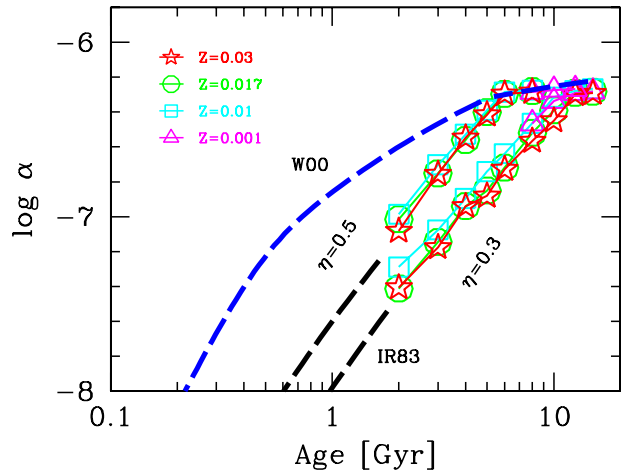,width=\hsize,clip=}
}
\caption{Theoretical time evolution of the luminosity-specific PN number
for SSP models of Table~\ref{alpha} (both for $\eta = 0.3$ and
0.5 and the different metallicity values, as labeled top left on the
plot) compared to the expected calibrations assuming the theoretical
IFMR of \citeauthor*{iben} and the empirical one from
\citeauthor*{weidemann}.  }
\label{alpha_ssp}
\end{figure}

\begin{table*}
\begin{minipage}{0.9\hsize}
\caption{Luminosity-specific PN number for Salpeter SSPs$^{(a)}$}
\begin{tabular}{rrrrrrrrrrrrrr}
\hline
  {Age~}  & \multicolumn{11} {c}{\hrulefill ~~Metallicity [Z]~~ \hrulefill} & & \\
   {[Gyr]}     & {0.001} & {0.01} & {0.017} & {0.03}  & IR83 & & {0.001} & {0.01} & {0.017} & {0.03} & IR83 & & W00\\
         & \multicolumn{5} {c}{\hrulefill$~~ \eta = 0.3 ~~$ \hrulefill} & \hrulefill & \multicolumn{5} {c}{\hrulefill$~~ \eta = 0.5 ~~$ \hrulefill} \\ 
\hline
  0.1 &  ...   &  ...	&  ...    &  ...   & --10.33 &   &  ...   &  ...   &  ...   &  ...   & --9.71&   & --8.71 \\
  0.5 &  ...   &  ...	&  ...    &  ...   & --8.57  &   &  ...   &  ...   &  ...   &  ...   & --8.16&   & --7.22 \\
  1.0 &  ...   &  ...	&  ...    &  ...   & --7.98  &   &  ...   &  ...   &  ...   &  ...   & --7.60&   & --6.86 \\
  2.0 &  ...   & --7.29 &  --7.41 & --7.41 & --7.45  &   &  ...   & --6.99 & --7.02 & --7.08 & --7.11&   & --6.60 \\
  3.0 &  ...   & --7.08 &  --7.15 & --7.18 & --7.17  &   &  ...   & --6.70 & --6.75 & --6.76 & --6.85&   & --6.47 \\
  4.0 &  ...   & --6.90 &  --6.94 & --6.94 & --6.98  &   &  ...   & --6.53 & --6.56 & --6.55 & --6.68&   & --6.38 \\
  5.0 &  ...   & --6.74 &  --6.85 & --6.88 & --6.84  &   &  ...   & --6.39 & --6.40 & --6.42 & --6.54&   & --6.32 \\  
  6.0 &  ...   & --6.64 &  --6.72 & --6.73 & --6.73  &   &  ...   & --6.29 & --6.30 & --6.30 & --6.45&   & --6.30 \\
  8.0 & --6.47 & --6.46 &  --6.52 & --6.57 & --6.56  &   & --6.29 & $\le$--6.28 & $\le$--6.28 & $\le$--6.29 & --6.28&   & --6.27 \\
 10.0 & --6.34 & --6.31 &  --6.39 & --6.45 & --6.44  &   & $\le$--6.27 & \multicolumn{3}{l}{\vbox{\hbox{------------------------------------}\hbox{$\mid$}} $\mid$} & --6.27 &  & --6.27 \\  
 12.5 & --6.29 & --6.28 &  --6.29 & --6.30 & --6.31  &   & \multicolumn{4}{l}{\vbox{\hbox{---------------}\hbox{$\mid$}} \hfill {\large{{no PNe}$^{(b)}$}} \hfill $\mid$} & --6.27 &  & --6.27  \\
 15.0 & $\le$--6.28 & $\le$--6.27 & $\le$--6.28 & $\le$--6.29 & --6.27  &   & \multicolumn{1}{l}{$\mid$} & \multicolumn{3}{r}{$\mid$} & $\le$--6.27 &  & $\le$--6.27 \\
\hline
\end{tabular}

$^{(a)}$ The listed quantity is $\log \alpha = \log N_{\rm PN} - \log (L_{\rm SSP}/L_\odot)$.\\
$^{(b)}$ Planetary nebulae not expected to form for these age/metallicity combinations as $M_{\rm PAGB} \le 0.52\,M_\odot$.
\label{alpha}
\end{minipage}
\end{table*}

The theoretical luminosity-specific PN number for \citeauthor*{b89} SSPs of different age, 
metallicity and mass-loss parameter is summarized in Fig.~\ref{mcr} and \ref{alpha_ssp},
by matching the IFMR prescriptions according both to case {\it (a)} and {\it (b)}, as in previous
discussion. As expected, Fig.~\ref{mcr} shows that $M_{\rm core}$ is
indeed the leading parameter constraining $\alpha$; note however the
relevant difference between the \citeauthor*{iben} and
\citeauthor*{weidemann} models, with the latter reaching a fixed core
mass at younger age, which corresponds to a brighter SSP
luminosity. Compared to \citeauthor*{iben}, therefore, the
\citeauthor*{weidemann} IFMR predicts in general a lower value of
$\alpha$ {\it for fixed value of $M_{\rm core}$}.

In this framework, the role of other SSP distinctive parameters, like metallicity and 
mass loss, only enters at a later crucial stage of the analysis, when tuning 
up the clock that links $M_{\rm core}$ to SSP age.
This is displayed in Fig.~\ref{alpha_ssp}, where the theoretical evolution of $\alpha$ 
is traced for SSPs spanning the full range of metal abundances and Reimers $\eta$ parameter.
Again, the \citeauthor*{iben} $\eta = 0.3$ and 0.5 cases and the \citeauthor*{weidemann} 
IFMR are compared in the figure. When the variation of $\alpha$ is predicted as a 
function of the SSP age, the \citeauthor*{weidemann} model leads to a systematically 
higher PN number per unit SSP luminosity, compared to the \citeauthor*{iben} case,
due to a lower value of PAGB core mass assumed.

A summary of our calculations is reported in Table~\ref{alpha}.
A Salpeter IMF was adopted in the models, but a change in the
power-law index with respect to the canonical value of $s = 2.35$ can
easily be accounted for, following eq.~(\ref{eq:4}):
\begin{equation}
\Delta \log \alpha = \log \left({\alpha_{\rm s} \over \alpha_{\rm Sal}}\right) = 
\log \left({{\cal B}_{\rm s} \over {\cal B}_{\rm Sal}}\right) = 
\log \left({{\cal L}_{\rm s} \over {\cal L}_{\rm Sal}}\right).
\end{equation}
This implies that, for a giant-dominated SSP ($s = 1.35$), $\log \alpha$
is $\sim 0.04$~dex higher, and the opposite happens for a
dwarf-dominated SSP ($s = 3.35$), for which $\log \alpha$ is $\sim
0.10$~dex lower than for the Salpeter case (see Fig.~\ref{bb}).

The effect of enhanced mass loss on the Fig.~\ref{alpha_ssp} models can be 
quantified in roughly $\Delta \log \alpha \simeq +0.4$~dex for a change of $\eta$ 
from 0.3 to 0.5 and fixed SSP age. This is a consequence of a lower core mass and a 
correspondingly slower PAGB evolution (see footnote~\ref{nota1}), recalling that 
$\Delta \log \alpha = \Delta \log \tau_{\rm PAGB}$.

A different value for the evaporation timescale directly reflects on
$\alpha_{\rm max}$. From eq.~(\ref{eq:amax}), as $\Delta \log
\alpha_{\rm max} = \Delta \log \tau_{\rm dyn}$, one has for instance that
$\Delta \log \alpha_{\rm max} \simeq -0.18$~dex when $\tau_{\rm dyn}$ is 
decreased, say, to 20\,000~yr.

\begin{table*}
\begin{minipage}{\hsize}
\caption{The luminosity-specific PN number for template galaxy models$^{(a)}$}
\begin{tabular}{rrrrrrrrrrrrrrrr}
\hline
 {Age~}  & & M$_{\rm bol}$ & B--V & $\log \alpha_{03}$ & $\log \alpha_W$ & & M$_{\rm bol}$ & B--V & $\log \alpha_{03}$ & $\log \alpha_W$ & & M$_{\rm bol}$ & B--V & $\log \alpha_{03}$ & $\log \alpha_W$ \\
 {[Gyr]} & & & & & & & & & & & &  \\
\hline
  & & \multicolumn{4}{c}{\hrulefill~~E~~\hrulefill} & & \multicolumn{4}{c}{\hrulefill~~Sa~~\hrulefill} & & \multicolumn{4}{c}{\hrulefill~~Sb~~\hrulefill} \\  
 & & & & & & & & & & & & \\
  1.0~~   & & --23.10 & 0.66 & --7.84 &  --6.87 &  & --23.22 &  0.58 & --7.94 &  --6.98 & &  --23.11 &  0.55 & --7.99 &  --7.02 \\    
  2.0~~   & & --22.48 & 0.72 & --7.40 &  --6.56 &  & --22.67 &  0.63 & --7.51 &  --6.67 & &  --22.66 &  0.58 & --7.59 &  --6.74 \\
  3.0~~   & & --22.13 & 0.76 & --7.16 &  --6.43 &  & --22.36 &  0.65 & --7.27 &  --6.54 & &  --22.42 &  0.59 & --7.36 &  --6.63 \\
  4.0~~   & & --21.89 & 0.79 & --6.98 &  --6.36 &  & --22.14 &  0.67 & --7.09 &  --6.47 & &  --22.25 &  0.60 & --7.20 &  --6.57 \\
  5.0~~   & & --21.70 & 0.81 & --6.84 &  --6.32 &  & --21.97 &  0.69 & --6.95 &  --6.43 & &  --22.13 &  0.61 & --7.08 &  --6.54 \\
  6.0~~   & & --21.54 & 0.83 & --6.73 &  --6.30 &  & --21.84 &  0.70 & --6.84 &  --6.41 & &  --22.03 &  0.62 & --6.97 &  --6.52 \\ 
  8.0~~   & & --21.30 & 0.86 & --6.55 &  --6.28 &  & --21.62 &  0.72 & --6.66 &  --6.38 & &  --21.88 &  0.63 & --6.81 &  --6.49 \\
 10.0~~   & & --21.11 & 0.88 & --6.41 &  --6.27 &  & --21.45 &  0.74 & --6.53 &  --6.37 & &  --21.76 &  0.63 & --6.68 &  --6.48 \\
 12.5~~   & & --20.92 & 0.90 & --6.29 &  --6.28 &  & --21.29 &  0.75 & --6.42 &  --6.37 & &  --21.65 &  0.64 & --6.58 &  --6.48 \\
 15.0~~   & & --20.76 & 0.92 & --6.29 &  --6.29 &  & --21.15 &  0.76 & --6.41 &  --6.38 & &  --21.57 &  0.65 & --6.57 &  --6.49 \\
\hline
  & & \multicolumn{4}{c}{\hrulefill~~Sc~~\hrulefill} & & \multicolumn{4}{c}{\hrulefill~~Sd~~\hrulefill} & & \multicolumn{4}{c}{\hrulefill~~Im~~\hrulefill} \\  
 & & ~~ & & & & & & & & & \\
  1.0~~   & & --22.80 & 0.55 & --7.99 &  --7.03 &  & --22.12 &  0.51 & --8.06 &  --7.09 & &  --20.29 &  0.30 & --9.03 &  --7.73 \\
  2.0~~   & & --22.47 & 0.55 & --7.65 &  --6.80 &  & --22.05 &  0.48 & --7.80 &  --6.94 & &  --21.06 &  0.34 & --8.57 &  --7.55 \\
  3.0~~   & & --22.34 & 0.54 & --7.46 &  --6.72 &  & --22.11 &  0.46 & --7.66 &  --6.91 & &  --21.49 &  0.36 & --8.31 &  --7.41 \\
  4.0~~   & & --22.26 & 0.54 & --7.33 &  --6.68 &  & --22.18 &  0.46 & --7.56 &  --6.89 & &  --21.80 &  0.38 & --8.13 &  --7.30 \\
  5.0~~   & & --22.22 & 0.54 & --7.22 &  --6.66 &  & --22.25 &  0.46 & --7.48 &  --6.88 & &  --22.04 &  0.39 & --7.98 &  --7.21 \\
  6.0~~   & & --22.19 & 0.54 & --7.14 &  --6.65 &  & --22.32 &  0.46 & --7.41 &  --6.87 & &  --22.24 &  0.40 & --7.86 &  --7.15 \\ 
  8.0~~   & & --22.15 & 0.54 & --7.00 &  --6.64 &  & --22.44 &  0.47 & --7.29 &  --6.84 & &  --22.55 &  0.42 & --7.67 &  --7.05 \\
 10.0~~   & & --22.13 & 0.55 & --6.90 &  --6.63 &  & --22.54 &  0.48 & --7.19 &  --6.82 & &  --22.79 &  0.43 & --7.53 &  --6.99 \\
 12.5~~   & & --22.12 & 0.55 & --6.81 &  --6.63 &  & --22.64 &  0.49 & --7.10 &  --6.79 & &  --23.03 &  0.45 & --7.39 &  --6.93 \\
 15.0~~   & & --22.11 & 0.56 & --6.79 &  --6.63 &  & --22.73 &  0.49 & --7.05 &  --6.77 & &  --23.23 &  0.46 & --7.29 &  --6.88 \\
\hline
\end{tabular}

$^{(a)}$ Models are for a Salpeter IMF;\\
$\log \alpha_{03}$ = luminosity-specific PN number assuming a theoretical IFMR with a fixed Reimers mass loss parameter $\eta = 0.3$;\\
$\log \alpha_W$ = luminosity-specific PN number assuming an empirical IFMR according to \citet{weidemann}.
 \label{temp}
\end{minipage}
\end{table*}

Given the extreme uncertainty of theory in quantitatively assessing
the PAGB transition time, in our SSP models we do {\it not}
explicitly account for the quick drop of luminosity-specific PN
number for the $0.52~M_\odot \lesssim M_{\rm core} \lesssim
0.55~M_\odot$ PAGB core mass range.  Operationally, as far as the PN
evolutionary regime is driven by $\tau_{\rm dyn}$, we neglect the
$\tau_{\rm tt}$ contribution and simply assume $\alpha = \alpha_{\rm
max},$\footnote{In this case, an upper limit is marked for $\alpha$ in
Table~\ref{alpha}.} while for $M_{\rm core} \le 0.52~M_\odot$ no PN
events are expected to occur at all and $\alpha = 0$. This crude
simplification has negligible consequences when modeling late-type
galaxies, while the evolution in the $0.52 \lesssim M_{\rm core}
\lesssim 0.55~M_\odot$ mass range becomes important for early-type
galaxy models, as we shall discuss in more detail in Sec.~\ref{ell}.

As a concluding remark, we must also recall that $\alpha$ vanishes for SSP
ages $t \lesssim 10^8$~yr, when the prevailing high-mass stars of
5-7~M$_\odot$ or higher override the PN phase and end up their
evolution as Supernovae.

\section{The luminosity-specific PN number in galaxies}
\label{galx}

The evolutionary properties of PNe in SSPs are the basis for extending
the analysis to a wider range of star formation histories, as it
happens in real galaxies. We use the template galaxy models developed
by \citet{b02,b05}, which ensure a self-consistent treatment of the PN
evolution and the photometric properties of the stellar populations in
the parent galaxy. In this framework, colors and morphological features along 
the Hubble sequence for early- and late-type systems were reproduced tracing
the individual luminosity contribution for
the bulge, disk and halo component \citep[see][for additional details]{b02,b05} .

Following Sec.~\ref{mc}, the PN evolution has been implemented in the models
in a semi-analytical way, by fitting the SSP data in Table~\ref{alpha}
for the $\eta = 0.3$ and 0.5 cases, and for the
\citeauthor*{weidemann} empirical IFMR. The whole data grid can be
fitted within a 6\% formal uncertainty in the value of $\alpha$ [i.e.\
$\sigma(\log \alpha) = \pm 0.025$~dex] over the age/metallicity range
by
\begin{equation}
\begin{array}[]{ll}
\log \alpha' = & (1.52 -0.07\,z)\,\log t_9 -0.07\,\log^2\left({3.4\,z}\right)- \\
               &  -0.1/t_9 +2.0\,(\eta-0.3) -7.80,
\end{array}
\label{pippo}
\end{equation}
where $t_9$ is the SSP age in Gyr, and $z = Z/Z_\odot$.
For solar metallicity, eq.~(\ref{pippo}) reproduces the \citeauthor*{iben}
calibration over the extended age range of Table~\ref{alpha}.
Similarly, a fit to the \citeauthor*{weidemann} relation is:
\begin{equation}
\log \alpha' = -0.6\,(\log t_9 - 1)^2 -6.27.
\end{equation}
In addition, we also assume that
\begin{equation}
\alpha = \left\{
\begin{array}[]{ll}
{\rm min}\,[\alpha', \alpha_{\rm max}], & \qquad {\rm for}~t_9 \ge 0.1 \\
0  & \qquad {\rm for}~t_9 < 0.1
\end{array}
\right.
\end{equation}
where $\alpha_{\rm max}$ is given by eq.~(\ref{eq:amax}).

The absolute number of PNe in a SSP of total (bolometric) luminosity
$L_{\rm SSP}$ simply becomes $N_{\rm PN}(\rm SSP) = \alpha\,L_{\rm SSP},$
and for a star-forming galaxy of age $t$ we write
\begin{equation}
{\cal N}_{\rm PN}(t)  = \int_0^t \alpha(\tau)\,L_{\rm SSP}(\tau)\ SFR(t-\tau)\ d\tau.
\label{eq:pngal}
\end{equation}

Therefore, the global luminosity-specific PN number for the galaxy
stellar population can be computed as
\begin{equation}
\alpha(t) =  {{\cal N}_{\rm PN}(t) \over L_{\rm gal}} = 
{ {\int_0^t \alpha(\tau)\,L_{\rm SSP}(\tau)\ SFR(t-\tau)\ d\tau} \over {\int_0^t L_{\rm SSP}(\tau)\ SFR(t-\tau)\ d\tau} }.
\label{eq:galx}
\end{equation}

\begin{figure*}
\centerline{
\psfig{file=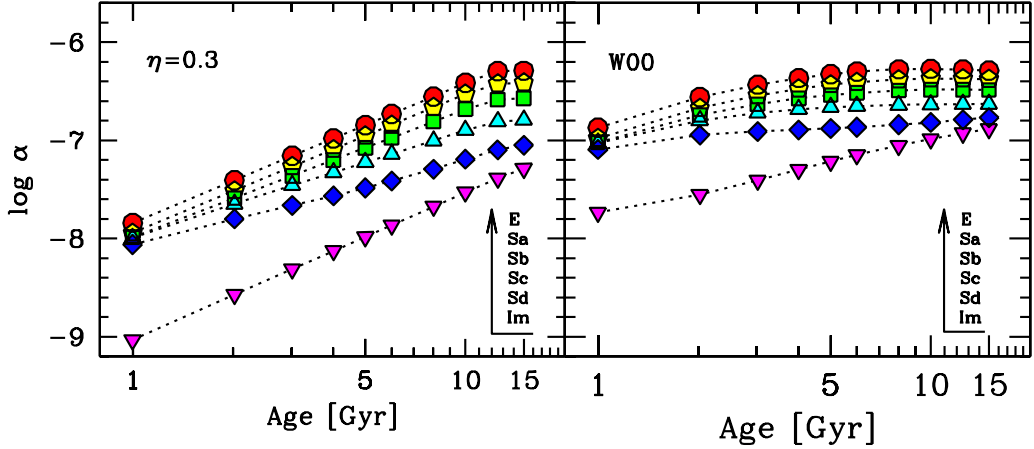,width=0.83\hsize,clip=}
}
\caption{Theoretical time evolution of the luminosity specific PN
density ($\alpha$) for the \citet{b05} template galaxy models along
the whole E-Sa-Sb-Sc-Sd-Im Hubble morphological sequence.  Models in
left panel assume an IFMR as from the standard mass loss parameter
$\eta = 0.3$, while those in the right panel rely on the empirical
relation from \citeauthor*{weidemann}.  Note, in the latter case, the
much shallower evolution of $\alpha$. In the two panels,
bulge-dominated spirals tend to approach the evolution of ellipticals
at early epochs due to the increasing bulge contribution to the global
galaxy luminosity.  }
\label{pngal}
\end{figure*}

The results for the whole Hubble sequence, from type E to Im, are
summarized in Fig.~\ref{pngal} and Table~\ref{temp} for the $\eta =
0.3$ case ($\alpha_{03}$ in the table) and the \citeauthor*{weidemann}
IFMR ($\alpha_W$).  In Table~\ref{temp} we provide also the integrated
$B-V$ color for the parent galaxy, and the absolute bolometric
luminosity of the system, assuming a total stellar mass of $M_{\rm
gal} = 10^{11}M_\odot$ at 15 Gyr \citep[see][for a detailed definition
of this quantity]{b05}.\footnote{The \citet{b05} galaxy models are computed with
a standard mass-loss parameter $\eta = 0.3$, and the integrated
magnitudes and colors reported in Table~\ref{temp} refer to this
case. The effect on galaxy $B-V$ by adopting the alternative
\citeauthor*{weidemann} IFMR can be estimated in $\Delta (B-V) \simeq
-0.02$~mag, that is with a little shift toward bluer colors \citep[see
e.g., the calibration from Fig.~28 in][]{b95}.  The value reported in
Table~\ref{temp} for $\log \alpha_W$ should also be increased by an
additional $\Delta \log \alpha_W \simeq +0.04$~dex if one takes into
account slightly fainter galaxies (some --10\% in bolometric
luminosity) in the \citeauthor*{weidemann} framework.\label{foot}}

The two evolutionary scenarios produce large differences in
$\alpha(t)$ for the Hubble morphological types: the
\citeauthor*{weidemann} model predicts larger PN populations and a
shallower time dependence for $\alpha$ than for the $\eta = 0.3$ case.
The effects are larger for late-type galaxy models, where the
impact of the different IFMR details on intermediate- and high-mass
stars is stronger.  Also, the nucleated late-type galaxies (types Sa
$\Rightarrow$ Sd) approach the evolution of ellipticals at early
epochs.  As discussed in some detail in \citet{b05}, this effect is
due to the prevailing photometric contribution of the bulge stellar
component, when $t \to 0$. 

The general prediction from these models is that {\it $\alpha$ is
expected to decrease in young and/or star-forming galaxies, compared
to more ``quiescent'' early-type systems} as a consequence of a
smaller population of PNe embedded in a higher galaxy luminosity per
unit mass (i.e.\ a lower $M/L$ ratio).

\subsection{Comparing with the observations: PN luminosity function 
and completeness corrections\label{comparison}}

The empirical evidence, from the 5007~\AA\ [O{\sc iii}] PN luminosity
distribution, indicates a nearly constant value for the bright cut-off
magnitude ($M^*$) of the PNLF.  This feature is actually found to be
nearly invariant with galaxy type and age and, as suggested by
\citet{jacoby89}, can be effectively used as a standard candle to
determine extragalactic distances.

When the emission-line fluxes are converted into equivalent V
magnitudes via the formula 
\begin{equation}
m_{\rm [O~III]}=-2.5\log F_{\rm [O~III]}-13.74
\end{equation}
\citep{jacoby89}, the PNLF takes the shape of a double-exponential
function \citep[][]{cia89} of the form
\begin{equation}
\log N(M) = 0.133\,M +\log [1-e^{3(M^{\ast}-M)}] +{\rm const},
\label{eq:F-PNLF}
\end{equation}
with the bright cut-off magnitude placed at $M^{\ast}$=$-4.47$~mag plus a little
metallicity correction, that scales with PN Oxygen abundance, such as
\begin{equation}
\Delta M^{\ast} = 0.928\,[O/H]^2 + 0.225\,[O/H]+0.014
\label{F-META}
\end{equation}
\citep{dopita92}. After correction, the inferred PN distances in external
galaxies are consistent within 0.1~dex with those obtained using the
Cepheids method for a large observed sample of galaxies and galaxy
types \citep{ciardullo}.

\begin{figure}
\centerline{
\psfig{file=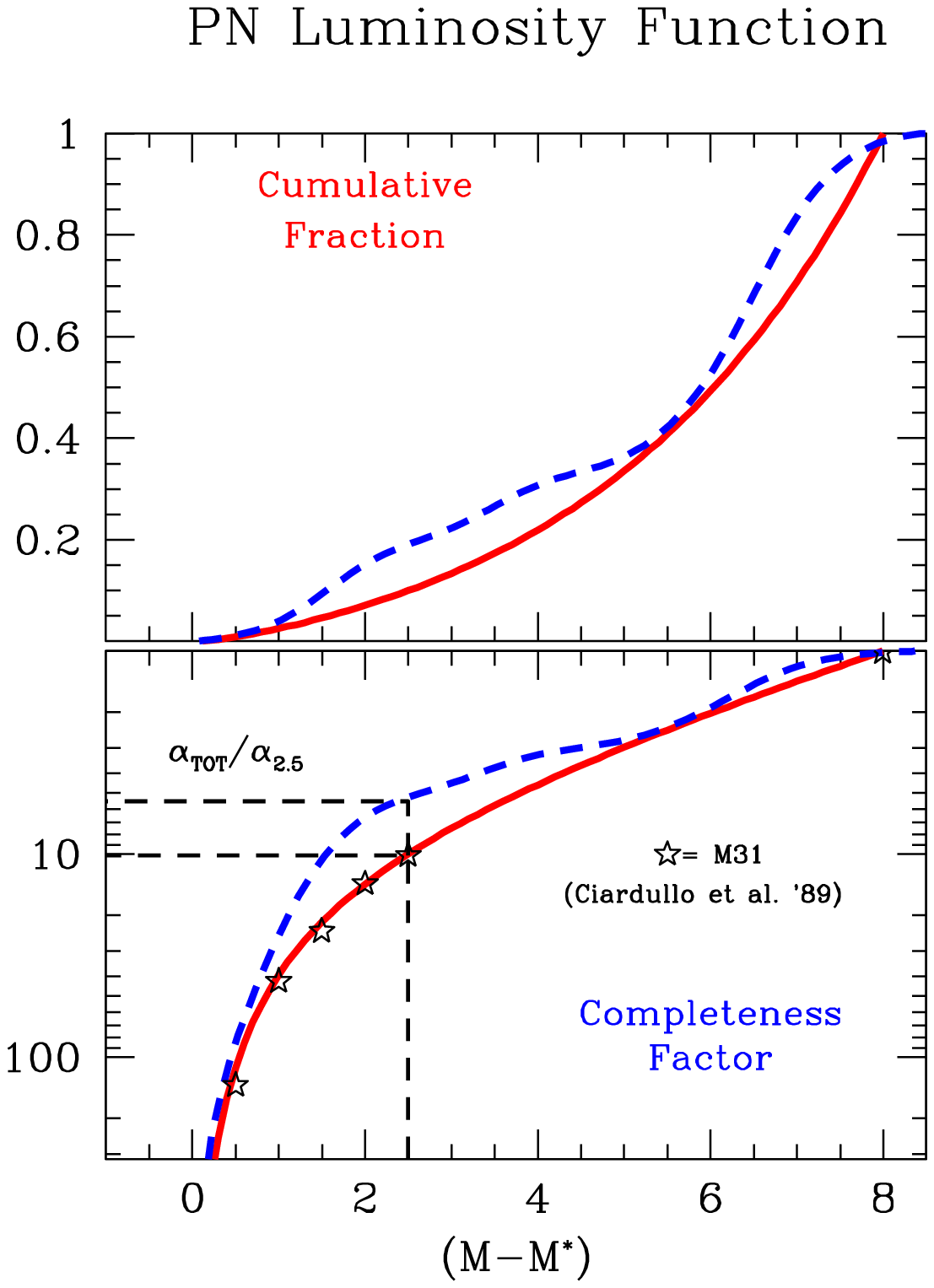,width=0.92\hsize,clip=}
}
\caption{{\it Upper panel:} the cumulative fraction of PNe in the different 
magnitude bins with respect to the luminosity-function bright cut-off ($M^*$)
for the double-exponential fit of the PNLF, as in eq.~(\ref{eq:F-PNLF})
(solid curve) and for the empirical SMC luminosity function
according to \citet{jacoby05} (dashed curve).
{\it Lower panel:} completeness factor ($CF = N_{\rm tot}/N_{\rm (M-M^*)}$) for
the same calibrations as in the upper panel. Also reported are the \citet{cia89}
data for M~31 (star markers), and the relevant correction factor for the
$\alpha_{2.5}$ parameter. For better convenience, data are also listed in
Table~\ref{lf}.
}
\label{pnlf}
\end{figure}

\begin{table}
\caption{The PN Luminosity Functions}
\begin{tabular}{crrrrr}
\hline
 M--M$^*$ & \multicolumn{2}{c}{\hrulefill~~Cumulative Fraction~~\hrulefill}  &  & \multicolumn{2}{c}{\hrulefill~~Completeness Factor~~\hrulefill} \\
 $[$mag$]$ & Standard &  SMC    &  & Standard  &  SMC  \\
\hline
 0.0 &     0 & 0     &  & $\infty$ & $\infty$ \\  
 0.5 & 0.009 & 0.013 &  & 113. & 78.3 \\  
 1.0 & 0.025 & 0.042 &  & 39.8 & 23.7 \\  
 1.5 & 0.046 & 0.097 &  & 21.6 & 10.3 \\  
 2.0 & 0.071 & 0.155 &  & 14.1 & 6.46 \\ 
 2.5 & 0.100 & 0.193 &  & 9.99 & 5.19 \\ 
 3.0 & 0.134 & 0.225 &  & 7.46 & 4.44 \\ 
 3.5 & 0.173 & 0.269 &  & 5.76 & 3.72 \\ 
 4.0 & 0.219 & 0.310 &  & 4.56 & 3.23 \\  
 4.5 & 0.273 & 0.336 &  & 3.66 & 2.98 \\ 
 5.0 & 0.336 & 0.366 &  & 2.98 & 2.73 \\ 
 5.5 & 0.409 & 0.427 &  & 2.45 & 2.34 \\ 
 6.0 & 0.494 & 0.535 &  & 2.03 & 1.87 \\ 
 6.5 & 0.593 & 0.692 &  & 1.69 & 1.44 \\ 
 7.0 & 0.708 & 0.845 &  & 1.41 & 1.18 \\ 
 7.5 & 0.843 & 0.941 &  & 1.19 & 1.06 \\ 
 8.0 &     1 &  1    &  & 1.00 & 1.00 \\  
\hline
\end{tabular}
\label{lf}
\end{table}

There is no theoretical explanation of this ``universal'' property of
the PN distribution, although it has been questioned \citep[e.g.][]{mendez} that 
it might depend on the sample size, being therefore 
a mere statistical effect. Furthermore, on the theoretical side, some change of $M^*$
with the age of the PN parent stellar population should be expected, 
its precise amount strongly depending, however, on the model assumptions
\citep[see especially][for the most striking results on this line]{marigo04}.

While the bright end of the PNLF has been studied in a large number of
galaxies, not much is known about the PNLF shape at fainter magnitudes, as
most Local Group (LG) surveys are complete down to 4 mag from the bright
cut-off luminosity \citep{cm}, and this limit becomes obviously brighter for more
distant systems. This implies a still large uncertainty in the
extrapolation from the {\it observed} number of PNe to the {\it whole} PN
population size, through eq.~({\ref{eq:F-PNLF}}). In particular, 
there is evidence of a dip in the PNLF of the SMC \citep{jdm} and M33
\citep{mag,cia04} at 4 and 2.5~mag, respectively, below the bright
cut-off (cf.\ Fig.~\ref{pnlf}, dashed curve in the bottom panel), but
this is not observed in the M31 bulge \citep{ciardullo}. The presence of
the dip might depend on the galaxy star formation history (through the
age distribution of PN progenitors), possibly witnessing the presence of 
a significant component of young stars \citep{cia04,marigo04}

According to \citet{cia89} and \citet{jacoby}, the faint-end tail of the
standard PNLF agrees with the theoretical luminosity function of \citet{henize}, in
which a PN is modelled as a uniformly expanding homogeneous gas
sphere ionized by a non-evolving central star. The number of nebulae in
each luminosity interval should then be proportional to the PN lifetime  
spent within that luminosity bin. For a global PN lifetime of
30\,000 yr, \citet{henize} predict that faintest PNe should locate  
about 8 magnitudes below $M^*$.

Deep observations in the SMC \citep{jdm,jacoby05}, reaching more than
6~mag below the bright cut-off, show a significant decline of the
number of PNe compared to what predicted by the double-exponential
formula.  Clearly, a more extended sample from other nearby galaxies
at comparable magnitude depth would be required to better assess this
important problem.  Unfortunately, in case of late-type galaxies, the
[O{\sc iii}] PN detection is not quite a simple task; due to the
ongoing star formation, only a very small fraction of the [O{\sc iii}]
emission-line sources in spirals and irregulars is represented by
genuine PNe, in most cases counts are affected by H{\sc ii} regions
and supernova remnants.  Identification of PNe in LG galaxies requires
additional constraints, such as point-like on-line emission combined
with a non-detection in the off-line continuum, and a stronger [O{\sc
iii}] emission compared to H$\alpha$ and/or [N{\sc ii}] narrow-band
luminosity.\footnote{A stronger [O{\sc iii}] emission vs.\ H$\alpha$
is found to be an excellent diagnostic tool for bright PNe
\citep{cia04}, while a reversed trend is likely to be expected for the
faintest nebulae \citep{mag}. See, for instance, \citet{arnaboldi} and
\citet{corradi} for further diagnostic plots in the emission-line
narrow-band color domain.}  In addition, disk regions can be heavily
obscured by dust, and the sample completeness with respect to the
parent stellar population light may also be affected.

A comparison of the \citet{jacoby05} updated SMC PNLF with the
standard double-exponential function, as in eq.~({\ref{eq:F-PNLF}}),
is proposed in Table~\ref{lf} and Fig.~\ref{pnlf}.  If the faint-end
tail of the PNLF indeed occurs at 8 magnitudes down from the bright
cut-off, then one can determine the fraction of PNe in the brightest
2.5 mag range, and define a parameter $\alpha_{2.5}$ as the
luminosity-specific number of PNe with $M-M^* \le
2.5$. Figure~\ref{pnlf} and Table~\ref{lf} show that such bright
nebulae represent a fraction between 10\% and 20\% of the total PN
population.

\begin{figure}
\centerline{
\psfig{file=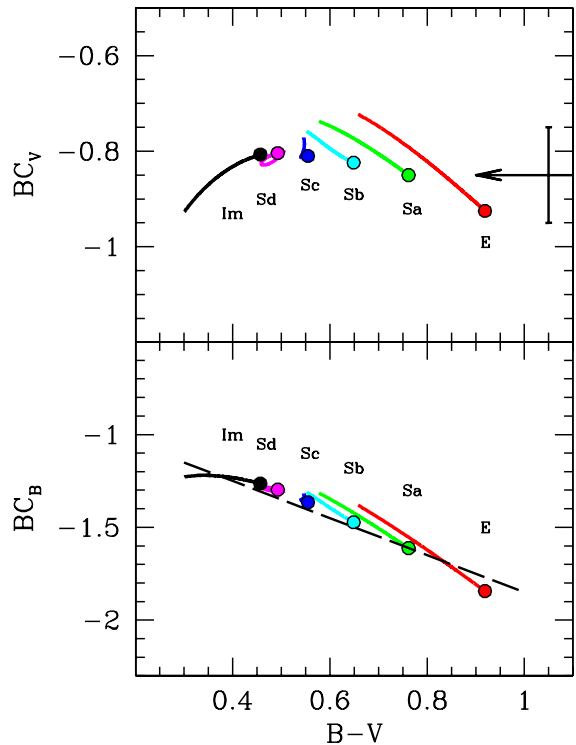,width=0.94\hsize,clip=}
}
\caption{The expected bolometric correction for theoretical template
galaxy models according to \citet{b05}. The models for different morphological
type span the age range from 1 to 15 Gyr (the latter limit being marked by the
solid dot on each curve). Bolometric correction refers to the $V$ 
({\it upper panel}) and $B$ band ({\it lower panel}). A  value of
$(Bol - V) = BC_V = -0.85$~mag can be taken as a representative correction 
for the whole galaxy types within a 10\% uncertainty, as shown by the arrow 
on the upper plot. This also translates into $BC_B \simeq -0.85 -(B-V)_{\rm gal}$
for the $B$-band correction, as displayed by the dashed line in the lower panel.
}

\label{bc}
\end{figure}

\begin{table*}
\begin{minipage}{0.93\hsize}
\scriptsize
\caption{The PN census in the Local Group galaxies$^{(a)}$}
\begin{tabular}{lllrlcrcrcrl}
\hline
Name      & Morph. & Distance & M$_B$ & $(B-V)_o$ & $[O/H]^{(b)}$ & Observed    & Completeness limit& \multicolumn{2}{c}{\hrulefill~~$N_{\rm tot}^{(c)}$~~\hrulefill}   & $\log \alpha \pm \sigma$ & Reference \\
          & Type   &  Mpc    &        &            &      & no.\ of PNe & M$_{\rm lim}$--M$^*$ & Standard & SMC        &  &  \\
\hline
M31 (all)   & Sb  & 0.76  & --21.55 &  $0.68 \pm 0.02$ & +0.1\,~ & $\sim$2700 &   &  \multicolumn{2}{c}{19000 $\pm$ 8000}	       &  & (1)\\
M31 (bulge) &     &	  & --18.01 &  $0.95 \pm 0.02^{(d)}$ &  &  94	    & 2.5 & 940 $\pm$ 97 & 497 $\pm$ 51 & $-6.94^{0.15}_{0.22}$ & (2) \\
\\
Milky Way   & Sbc & 0.01  & --20.80 &  $0.63^{(e)}$    & --0.2\,~ & $\sim$2000 &     &  \multicolumn{2}{c}{25000 $\pm$ 19000} & $-6.40^{0.25}_{0.63}$ & (3), (4) \\
\\
M\,33       & Scd & 0.80  & --18.74 &  $0.47 \pm 0.02$ & --0.5\,~ &  152      & 2.6 & 765 $\pm$ 85& 415 $\pm$ 46  & $-7.13^{0.14}_{0.22}$ &   \\
\\
LMC         & SBm & 0.050 & --17.93 &  $0.44 \pm 0.03$ & --0.52  & $\sim$1000 & $\sim7.0~~~~~$&  	 &            &  & (5)\\
            &     &	  &         &                  &      &$\sim$350  & $\sim5.0~~~~~$& 1040 $\pm$ 60& 960 $\pm$ 50 & $-6.57^{0.04}_{0.04}$  & (6)  \\
\\
SMC         & SBm & 0.060 & --16.24 &  $0.41 \pm 0.03$ & --0.84 &  105      & 6.0 &		& 167 $\pm$ 19    & $-6.67^{0.05}_{0.05}$  & (7),(6)\\
\\
NGC\,205    & E5  & 0.76  & --15.96 &  $0.82 \pm 0.05$ & --0.3\,~ &  35	    & 3.5 & 134 $\pm$ 28&  87 $\pm$ 18  & $-6.88^{0.15}_{0.22}$ &   \\
\\
M\,32       & E2  & 0.76  & --15.93 &  $0.88 \pm 0.01$ & --0.58 &  30	    & 2.4 & 186 $\pm$ 44&  97 $\pm$ 23  & $-6.77^{0.18}_{0.31}$ &   \\
\\
IC\,10      & Im  & 0.66  & --15.57 &  $0.58 \pm 0.05$ & --0.7\,~ &  16	    & 2.0 & 153 $\pm$ 46&  72 $\pm$ 22  & $-6.59^{0.20}_{0.40}$ &   \\
\\
NGC\,6822   & Im  & 0.50  & --15.22 &  $0.47 \pm 0.15^{(c)}$ & --0.62 &  17	    & 3.5 &  74 $\pm$ 21&  48 $\pm$ 13  & $-6.69^{0.16}_{0.27}$ &   \\
\\
NGC\,185    & E3  & 0.66  & --14.90 &  $0.73 \pm 0.01$ & --1.0\,~ &   5	    & 2.7 &  45 $\pm$ 20&  25 $\pm$ 11  & $-6.88^{0.22}_{0.45}$ &   \\
\\
NGC\,147    & E5  & 0.66  & --14.48 &  $0.78 \pm 0.05$ &    &   9	    & 3.9 &  29 $\pm$ 12&  20 $\pm$  8  & $-6.91^{0.19}_{0.34}$ &   \\
\\
Sex\,A      & Im  & 0.86  & --13.02 &  $0.37 \pm 0.08$ & --1.32 &  1	    & 1.9 &  16 $\pm$ 16&   7 $\pm$  7  & $-6.38^{0.30}_{\infty}$ &	\\
\\
Sex\,B      & Im  & 0.86  & --12.96 &  $0.51 \pm 0.03$ & --0.75 &  5	    & 3.0 &  37 $\pm$ 17&  22 $\pm$ 10  & $-6.10^{0.21}_{0.44}$ &   \\
\\
Leo\,A      & Im  & 0.69  & --11.36 &  $0.31 \pm 0.08$ & --1.51 &   1	    & 3.0 &	8 $\pm$  8&   5 $\pm$  5  & $-5.99^{0.30}_{\infty}$ &	\\
\hline
\end{tabular}
\smallskip

References: (1) \citet{merrett}; (2) \citet{cia89}; (3) \citet{jacoby}; (4) \citet{alloin}; (5) Reid \& Parker, private communication; (6) \citet{jacoby05}; (7) \citet{jdm}\\

$^{(a)}$ Galaxy morphological types and $(B-V)$ colors are from the RC3 catalog \citep{rc3}, distances and metallicities\\ 
$~~~~~~~\,$ from \citet{cm}, absolute B magnitudes, M$_B$, from \citet{karachentsev} rescaled to the \citet{cm} distance modulus. Both $(B-V)$ and $M_B$\\
$~~~~~~~\,$ quantities are reddening- and inclination-corrected according to the corresponding literature sources.\\
$^{(b)}$ Adopted PN Oxygen abundance, $[O/H] = \log (O/H) - \log (O/H)_\odot$. For the Sun, $12+\log(O/H)_\odot = 8.87$ \citep{grevesse}.\\
$^{(c)}$ PN population size within 8 mag from the PNLF bright-end tail, inferred according to
two different extrapolation methods:\\
$~~~~~~~\,$ Standard: empirical PNLF from \citet{jacoby89}, as in Table~\ref{lf} (columns 2 and 4), corrected for metallicity after \citet{dopita92}\\
$~~~~~~~\,$ SMC: using the \citet{jacoby05} observed SMC PNLF (see columns 3 and 5 in Table~\ref{lf}), and correcting for metallicity as for the ``Standard'' case.\\
$^{(d)}$ Data from HyperLeda Lyon.\\
$^{(e)}$ The $(B-V)$ estimate for the Milky Way is from the \citet{robin} synthetic model, as quoted by \citet{boissier}.
\label{census}
\end{minipage}
\end{table*}

As we will see in the following sections, the definition of
$\alpha_{2.5}$, which includes only the brighter PNe, makes the
comparison with observations in external galaxies easier. For our
discussion, we assume a simple relation such as $\alpha = 10 \times
\alpha_{2.5}$, according to the standard PNLF (column 4 in
Table~\ref{lf}). This normalization is correct if the
double-exponential PNLF formula applies, and there are no PNe 8~mag
fainter than the bright cut-off.

\subsubsection{Galaxy bolometric correction}
\label{bolcol}

To compare observations and model predictions for the
luminosity-specific PN number in external galaxies we need to convert
the monochromatic galaxy luminosity to bolometric.  One can
use the Johnson $V$ or $B$ magnitudes for galaxy photometry, corrected
for the distance modulus such as to match the absolute scale, and
usually complemented with a (rough) estimate of the color excess,
$E(B-V)$, to account for Galactic reddening.

Given the (reddening-corrected) values of $M_B$ and $M_V$, the
required transformations to provide $L_{\rm gal}$ in
eq.~(\ref{eq:galx}) are the following:
\begin{equation}
L_{\rm gal} = \left\{
\begin{array}[]{l}
10^{-0.4\,(M_B - 5.41)}\,10^{-0.4\,(BC_B +0.69)} \\
10^{-0.4\,(M_V - 4.79)}\,10^{-0.4\,(BC_V +0.07)}, 
\end{array}
\right.
\label{eq:bolcol}
\end{equation}
where $BC_B$ and $BC_V$ are the bolometric corrections to the $B$ and
$V$ band, respectively; according to \citeauthor*{b89}, in our
notation the Sun has $M^{\rm bol}_\odot = +4.72$~mag, $BC^V_\odot =
-0.07$~mag, and $BC^B_\odot = -0.69$~mag.  A direct estimate of the
bolometric correction from the galaxy observations is not a
straightforward task, and the alternative way is to rely on models.

Figure~\ref{bc} shows the trend of the $B$ and $V$ bolometric
correction for the \citet{b05} galaxy templates over the age range
from 1 to 15~Gyr.  One sees that a simple and convenient solution,
within a 10\% internal accuracy (i.e.\ $\pm 0.1$~mag) in the
transformation, can be found for the $V$-band correction, assuming a
fixed value $BC_V = (Bol-V) = -0.85$~mag all over the relevant range
of galaxy types and age.  A suitable estimate for the $B$-correction
is $BC_B = (Bol-V)-(B-V) = -0.85 -(B-V)$~mag.

\begin{figure*}
\centerline{
\psfig{file=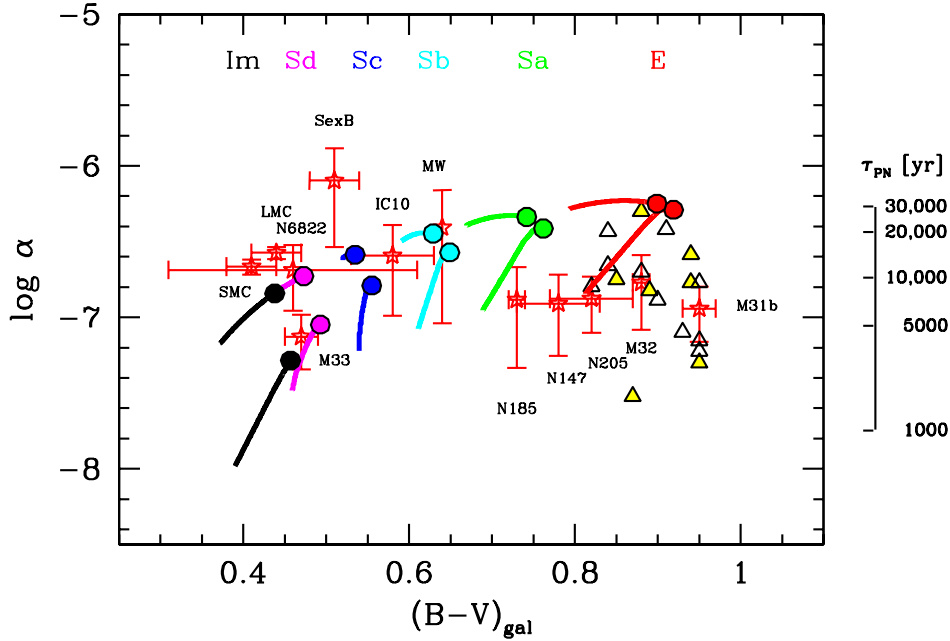,width=0.73\hsize,clip=}
}
\caption{A comprehensive overview of the luminosity-specific PN number
in LG galaxies (star markers) and external ellipticals from Table~\ref{earlygal}
(solid triangles) and Table~\ref{e-sample} (open triangles). PN data for local galaxies are
from Table~\ref{census}, and are based on the ``Local Group Census
Project'' of \citet{corradi}.  Also superposed on the plot, there are the
\citet{b05} template galaxy models, as summarized in Table~\ref{temp}.
Galaxy evolution is tracked by models along the whole E-Sa-Sb-Sc-Sd-Im
Hubble morphological sequence from 5 to 15 Gyr, with the latter limit
marked by the big solid dots. Two model sequences are reported on the
plot assuming an IFMR as from the standard case of a Reimers mass loss
parameter $\eta = 0.3$ (lower sequence), and from the empirical
relation of \citeauthor*{weidemann} (upper sequence).  For the
\citeauthor*{weidemann} models, the relevant data of Table~\ref{temp}
have been corrected by $\Delta (B-V) = -0.02$ mag and $\log \alpha =
\log \alpha_W + 0.04$, according to the arguments of
footnote~(\ref{foot}). An indicative estimate of the mean representative 
PN lifetime (in years) is sketched on the right scale, according 
to eq.~(\ref{eq:stima}).}
\label{templates}
\end{figure*}

\section{Comparison with observed PN populations}
\label{obs}

\subsection{The Local Group PN census}
\label{lgcensus}

The LG galaxies have been extensively searched for PNe since  the
discovery of five PNe in M31 by \citet{baade55} . Recent surveys
including all LG galaxies with $\log(L_B/L_\odot)>6.7$, provide
a comprehensive view of LG PN population which can be used as a first
comparison for our population synthesis models.  An updated
list of the number of PNe known in the LG can be found in the work of
\citet{cm}.  While PNe have been found in 20 LG galaxies, the survey
completeness and photometric accuracy are not good enough to allow a
confident estimate of their global population in all of them.

In Table~\ref{census} we therefore summarize the updated information
for the PN census only for those LG galaxies whose survey completeness
limit (and the number of PNe within it) is known.  Many of the data
come from the so-called Local Group Census project \citep[see e.g.][
and references therein]{corradi}; other sources are indicated in the
table.

From these data we have estimated the total PN population size of
these galaxies ($N_{\rm tot}$ in Table~\ref{census}) in two ways, to
account for the uncertainty in our knowledge of the shape of the PNLF
at faint magnitudes. First, for each galaxy the magnitude difference
between the completeness limit of the survey and the expected apparent
magnitude of the PNLF cut-off, $m^\star$, was computed. The value of
$m^\star$ is determined assuming $M^{\ast} = -4.47$, the metallicity
correction in eq.~(\ref{F-META}) with the value of [O/H] as reported
in column 6 of the table, and the distances as in \citet{cm} (listed
again in column 3 here). The observed number of PNe within the
completeness limit has then been extrapolated 8~mag down the PNLF
cut-off using the empirical formula in eq.~(\ref{eq:F-PNLF}), as
reported in Table~\ref{lf}; resulting figures are listed in column 9
of Table~\ref{census}. The total PN population size was also estimated
rescaling the {\it observed} PNLF of the SMC \citep[][see column 5 of
Table~\ref{lf}]{jdm}, which is complete 6~mag down the cut-off,
and assuming a $\sim 50$~\% incompleteness for the next magnitude bin,
consistent with recent deeper observations \citep{jacoby05}. The
corresponding total PN population estimated for the LG galaxies is
indicated in column 10 of Table~\ref{census}.

For the SMC, the adopted PN population is the number estimated by
\cite{jdm}, plus some 20\% to include the faintest PNe as suggested by
recent observations \citep{jacoby05}. The LMC population size is
determined using approximate figures of the observed number of PNe and
the depth of the discovery surveys, according to \citet{jacoby05}.
For M~31, data for a relatively dust-free region in the bulge
\citep{cia89} were adopted; the local number of PNe was then rescaled
to the whole galaxy luminosity. The latter value represents an
estimate of the global population, and the derived value of $\alpha$
reported in Table~\ref{census} is in fact representative of the bulge
evolutionary environment.  Finally, the total PN population for the
Milky Way in Table~\ref{census} is a conservative estimate from the
data by \citet{jacoby} and \citet{alloin}.

The difference between the total PN population based either on the
``standard'' or the ``SMC'' PNLF is of a factor two or smaller. The
derived value of $\alpha$ is reported in column 11 of
Table~\ref{census}, assuming for LG galaxies the $B-V$ colors from the
RC3 catalog \citep{rc3} and the absolute $M_B$ magnitudes from
\citet{karachentsev}, converted to bolometric according to
eq.~(\ref{eq:bolcol}).\footnote{To be consistent with the \citet{cm}
survey and completeness correction, the \citet{karachentsev} absolute
$M_B$ magnitudes have been slightly rescaled to the \citet{cm} adopted
galaxy distances, as reported in Table~\ref{census}. Note, in
addition, that both $M_B$ and $(B-V)$ are reddening- and
inclination-corrected values according to the original data sources.}
The quoted error bars in the Table are conservative estimates of the
uncertainty of $N_{\rm tot}$, which account for the difference between 
the ``standard'' and ``SMC'' PNLF extrapolated values.

\begin{table*}
\begin{minipage}{0.85\hsize}
\caption{Recent additions to PN census in early-type galaxies}
\begin{tabular}{lcccccl}
\hline
Name      & Morph. &  Observed    & Comp. limit& $N_{\rm tot}^{(b)}$ & $\log \alpha$ & Ref. \\
          & Type$^{(a)}$ &  no.\ of PNe & M--M$^*$ &  &	    &  \\
\hline
NGC~1316 & S0 & ~~43  & 1.0 & 1720 $\pm$ 262 & $-7.50\pm 0.07$ & \citet{arnaboldi98} \\  
\\
NGC~1344 & E5 & 197  & 1.0 & 2300 $\pm$ 47 & $-6.75 \pm 0.02$ & \citet{teodorescu} \\
\\
NGC~1399 & E1 & ~~37  & 1.0 & 1480 $\pm$ 243 & $-7.30\pm 0.07$ & \citet{arnaboldi94} \\  
\\
NGC~3115 & S0 & ~~61  & 1.0 & 2440 $\pm$ 312  & $-6.59\pm 0.05$ & \citet{ciardullo} \\
\\
NGC~3379 & E1 & 109 & 2.0 & 1535 $\pm$ 104 & $-6.77\pm 0.03$ & \citet{romanw03} \\
\\
NGC~4697 & E6 & 535 & 2.5 & 3500 $\pm$ 231 & $-6.82\pm 0.03$ & \citet{mendez01} \\
\\
NGC~5128 & S0 & 431 & 2.5 & 4300 $\pm$ 207 & $-6.30\pm 0.02$ & \citet{hui} \\
\hline
\end{tabular}
\label{earlygal}
\smallskip

$^{(a)}$ From the RC3 catalog \citep{rc3},\\
$^{(b)}$ PN population size within 8 mag from the PNLF bright-end cut-off, inferred according to standard\\
$~~~~~~$ PNLF, as in Table~\ref{lf} (columns 2 and 4).
\end{minipage}
\end{table*}

The LG galaxy sample provides the natural benchmark to test our
theoretical models over a range of evolutionary environments and star
formation histories. Our predictions for the \citet{b05} template
galaxy models of Table~\ref{temp} are compared in Fig.~\ref{templates}
with the LG data of Table~\ref{census} for a relatively old age range,
from 5 to 15 Gyr, and mass loss scenarios ($\eta = 0.3$ and
\citeauthor*{weidemann}).

The remarkable feature of the $\log \alpha$ plot, for the observed
galaxies representative of the whole Hubble morphological sequence,
is the fairly constant PN rate per unit galaxy luminosity. Data
support an average rate between 1 and 6~PNe per
$10^7$~L$_{\odot}$. Such a value is related to a narrow range of PAGB
stellar core mass, according to the calibration of Fig.~\ref{mcr},
with $M_{\rm core}$ less than 0.60-0.65~$M_\odot$.

There are at least three important consequences of this sharp mass
distribution: {\it i)} the mass-loss scenario supported by the
observations better agrees with the \citeauthor*{weidemann} IFMR,
which implies a stronger mass loss for intermediate and high-mass
stars compared to the standard scenario ($\eta \simeq 0.3$-0.5) for
Pop II stars as in Galactic globular clusters; {\it ii)} according to
Sec.~\ref{fc}, for a consistent fraction of the PN population in 
LG galaxies the inferred lifetime is constrained by the dynamical
timescale of nebula evaporation rather than the stellar core mass
evolution; {\it iii)} the latter evidence also supports a small
dependence of $\alpha$ with time and distance (see right panel of
Fig.~\ref{pngal}) pointing to a relatively ``universal'' shape for the
PNLF (but see, however, Sec.~\ref{diagnostic} for an important warning
in this regard).

\subsection{PN surveys in external galaxies}

As discussed in Sec.~\ref{comparison}, a complete survey of the PN
population in late-type galaxies is often plagued by spurious
detections caused by H{\sc ii} regions, SN remnants and, for galaxies
at distances larger than 10 Mpc, by Ly$\alpha$ background galaxies at $z
\simeq 3.13$, \citep[see e.g.][]{cia02, arnaboldi04}.  

Current surveys only investigate the brightest magnitude bin
of the PNLF, to measure the bright cut-off magnitude $M^*$
and lead therefrom to galaxy distance \citep[see, e.g.][for a recent
survey of six S0 and active, mostly Seyfert 2, spirals]{ciardullo}.
The best samples of PNe in late-type systems still remain
the ones for M31 and M33 \citep[][]{merrett,mag,cia04}.

\begin{table*}
\begin{minipage}{0.7\hsize}
\caption{The updated PN sample for local and distant early-type
galaxies$^{(a)}$}
\begin{tabular}{lcccccll}
\hline
  NGC & $\log \sigma$ & $(B-V)_o$ & Mg$_2$ & $(1550-V)_o$ & $\log \alpha^{(b)}$ & Ref. & Notes\\
       &  [km\,s$^{-1}$] &  [mag] &  [mag]  &  [mag]  &                 &         &  \\
\hline
~~205         & 1.61 &  0.82 & 0.071 & 1.19 & --6.88 & (1) & star forming\\ 
~~221         & 1.90 &  0.88 & 0.198 & 4.50 & --6.77 & (1) & M32\\
~~224         & 2.27 &  0.95 & 0.324 & 3.51 & --6.94 & (1) & M31 (bulge only) \\
 1316$^{(c)}$ & 2.38 &  0.87 & 0.265 & ~~~\,5.0~$^{(d)}$ & --7.50 & (2) & For A - merger\\
 1344         & 2.22 &  0.85 & 0.267 &      & --6.75 & (2) & \\
 1399         & 2.52 &  0.95 & 0.357 & 2.05 & --7.30 & (2) & \\
 3031$^{(c)}$ & 2.23 &  0.82 & 0.295 &      & --6.80 & (1) & \\
 3115         & 2.45 &  0.94 & 0.309 & 3.43 & --6.59 & (2) & \\
 3377$^{(c)}$ & 2.16 &  0.84 & 0.273 &      & --6.43 & (1) & \\
 3379         & 2.33 &  0.94 & 0.329 & 3.86 & --6.77 & (2) & \\
 3384$^{(c)}$ & 2.20 &  0.91 & 0.296 & ~~~\,3.9~$^{(d)}$ & --6.42 & (1) & \\
 4374         & 2.48 &  0.94 & 0.323 & 3.55 & --6.77 & (1) & \\
 4382         & 2.24 &  0.88 & 0.242 & 4.22 & --6.70 & (1) & \\
 4406         & 2.42 &  0.90 & 0.330 & 3.72 & --6.89 & (1) & \\
 4472         & 2.49 &  0.95 & 0.331 & 3.42 & --7.16 & (1) & \\
 4486         & 2.60 &  0.93 & 0.303 & 2.04 & --7.10 & (1) & \\
 4594$^{(c)}$ & 2.41 &  0.84 & 0.340 &      & --6.66 & (1) & \\
 4649         & 2.56 &  0.95 & 0.360 & 2.24 & --7.22 & (1) & \\
 4697         & 2.25 &  0.89 & 0.320 & 3.41 & --6.82 & (2) & \\
 5128$^{(c)}$ & 2.14 &  0.88 &       & $~~~\gg 5~?^{(e)}$ & --6.30 & (1) & Cen A - merger \\
\hline
\end{tabular}
\label{e-sample}
\smallskip

References for $\alpha$ estimates: (1) \citet{hui}; (2) this paper \\

$^{(a)}$ Stellar velocity dispersion $\log \sigma$, Mg$_2$ index and $(1550-V)_o$ color from \citet{burstein} unless\\
$~~~~~~~\,$ otherwise stated; reddening-corrected $(B-V)$ from the RC3 catalog \citep{rc3}; \\
$^{(b)}$ For the \citet{hui} data, $\log \alpha = \log \alpha_{2.5} +1.00$.\\
$^{(c)}$ $\log \sigma$ and Mg$_2$ from HyperLeda Lyon;\\
$^{(d)}$ $(1550-V)$ color estimated from the ultraviolet galaxy catalog of \citet{rifatto} and RC3/HyperLeda\\
$~~~~~~~\,$ dereddened total $V$ magnitude. Reddening correction is according to \citet{seaton}, assuming\\
$~~~~~~~\,$ a Galaxy extinction map from \citet{bh};\\
$^{(e)}$ Conservative estimate on the basis of the upper limit to the 1540~\AA\ galaxy flux, as reported by the NED\\
$~~~~~~~\,$ database.
\end{minipage}
\end{table*}

\subsubsection{PN samples in early-type galaxies}

The lack of active star formation and a negligible fraction of
residual gas in early-type galaxies make the [O{\sc iii}] PN detection
relatively straightforward across the whole galaxy body, excluding
the innermost regions, where the galaxy spectral continuum is too bright.

However, only low-mass ellipticals and dwarf spheroidals can be found
among the LG galaxy population and, with the exception of NGC~5128
(Cen A) at 3.5 Mpc, the nearest normal and/or giant ellipticals {\bf
are at 10~Mpc or further}.  At these distances, the typical [O{\sc
iii}] magnitudes are in the range 26-27~mag or fainter, and
spectroscopic-confirmed PN samples are usually limited to $M \lesssim
M^* +1.0$, with only few good cases observed in slitless
spectroscopy. Hopefully, larger PN samples in early-type galaxies will
soon be available with the dedicated PN.S spectrograph operating at
the William Herschel Telescope at La Palma \citep{douglas}.

A comprehensive collection of PN data for dwarfs, giant elliptical and
S0 galaxies in the LG, Leo group and the Virgo cluster can be found in
\citet{hui}, comparing original observations of NGC~5128 with the
value of $\alpha_{2.5}$ for a sample of 13 early-type galaxies plus
the bulge of M~31.  An updated census for some of these objects, plus
additional results from deeper surveys for a few more galaxies are
listed in Table~\ref{earlygal}.  In column~4 we report the total
number of detected PNe (at all magnitudes) together with the
completeness limit of the survey, ($M-M^*$), and the inferred number
of the global PN population on the sampled galaxy region (with its
Poissonian error estimate), by assuming a standard PNLF down to $M =
M^* + 8.0$~mag. The value of $\alpha$ is then computed by normalizing
to the sampled galaxy luminosity, according to the original data
sources in the literature.

For the data of Table~\ref{earlygal} and the \citet{hui} sample, 
we collected, in Table~\ref{e-sample}, supplementary dynamical
and photometric information from \citet{burstein}, the RC3, NED and
HyperLeda on-line databases. Two LG galaxies were also added to the
sample, including the dwarf satellite ellipticals of M~31 (i.e.\
NGC~205 and M~32), and the bulge data for M~31 itself. This whole
sample of early-type galaxies is displayed in Fig.~\ref{templates},
matching the LG data and the \citet{b05} template galaxy models.

As pointed out by \citet{hui}, in the plot of Fig.~\ref{templates}
one can notice a sharp decrease of the luminosity-specific PN number in
early-type galaxies compared with the theoretical predictions of our
E-galaxy model and the empirical estimate of $\alpha$ for
the Milky Way and other nearby spirals.  In addition, following
\citet{peimbert} and \citet{hui}, the figure also reports {\it
a clear trend of $\alpha$ with galaxy color, with a poorer PN population
(per unit bolometric luminosity) in redder ellipticals}. So
far, this trend has not received a satisfactory explanation.

\section{The PN population in ellipticals}
\label{ell}

From the definition of $\alpha$ in eq.~(\ref{eq:3}), a lower
value of the luminosity-specific PN number corresponds to a
shorter PN lifetime.  If $30\,000$~yr is a reasonable value for
$\tau_{\rm PN}$ in the most PN-rich galaxies, then
\begin{equation}
{\alpha\over\alpha_{\rm max}} \approx {\tau_{\rm PN} \over 30\,000}.
\label{eq:scale}
\end{equation}
For a low value of $\alpha$, like for instance in NGC~1316 
or NGC~1399, one leads therefore to a lifetime of 3000-5000~yr
for the galaxy PN population (see the right-scale  calibration in
Fig.~\ref{templates}).\footnote{Note that our definition of the PN lifetime
relates to the nebula visibility phase, and it is different from the
kinematic age of the system, as derived from the ratio between
absolute size of the main nebular shell and its expansion velocity
\citep[see, e.g.][for a discussion]{villaver}.} The value of
$\tau_{\rm PN}$ linked to the observed value of $\alpha$ in
eq.~(\ref{eq:scale}) must, however, be regarded as an average on the PAGB core
mass distribution, which cannot be determined observationally in
systems other than the Milky Way \citep[e.g.][]{zhang}.

As far as we restrain to standard stellar evolution theory, such a shorter (or even
a vanishing) PN lifetime can be obtained in three different (and to some extent 
mutually exclusive) ways, either  
\begin{enumerate}
\item by increasing the stellar core mass so that nuclear evolution speeds up 
(see footnote~\ref{nota1}),
\item by delaying the Hot-PAGB phase of the stellar core (for instance, by 
increasing $\tau_{\rm tt}$ in eq.~\ref{eq:alfa}) such as to let the nebula 
evaporate before it can be excited, or 
\item by fully inhibiting the AGB phase so that the PN event cannot take place, 
at least in a fraction of the galaxy stellar population. 
\end{enumerate}
Each of these different scenarios leaves a different ``signature'' in
the overall shape of the PNLF and its bright cut-off luminosity.
Definitely, case {\it (i)} above is the less favoured one in our
analysis, as it would predict {\it bluer} ellipticals to have a lower
value of $\alpha$, contrary to what we observe.  Likely, case {\it
(ii)} and {\it (iii)} better fit with an overall evolutionary scenario
dominated by low-mass (old?) PAGB stars, and they could be at work at
the same time (though to a different relative degree) among the PN
population of early-type galaxies, as the recent HST observations of
M~32 \citep{brown00} seem to support.  According to the arguments of
Sec.~\ref{casec}, the core mass range $0.52 \lesssim M_{\rm core}
\lesssim 0.55~M_\odot$ may be the preferred one to explain the case
{\it (ii)} scenario, while a more intense core mass depletion, leading
to $M_{\rm core} \lesssim 0.52~M_\odot$, would cause a fraction of
Post-HB stars to override the PN event, like in case {\it (iii)}.

\subsection{PN core mass and $M^*$ invariance}
\label{candle}

Within some still large theoretical uncertainties, stellar evolution
models \citep{ms97, marigo04} agree on the fact that the brightest PNe
are not always related to the most massive cores. Nonetheless, stars
of $M_{\rm i} \gtrsim 2\,M_\odot$, ending up their PAGB evolution with
$M_{\rm core} \gtrsim 0.7\,M_\odot$ (cf.\ Fig.~\ref{mfin}), are
required to generate the $M^*$ nebulae \citep[see, e.g.\ Fig.~10
in][]{marigo04}.

If our models for PN evolution are correct, then one predicts low-mass
nebulae ($M_{\rm core} \lesssim 0.65\,M_\odot$) to dominate the PNLF
of galaxies of different morphological type. In fact, such a claim is
even stronger for ellipticals, for which very low core-mass values
must be invoked on average for their old metal-rich stellar
populations. Therefore, it becomes harder in these galaxies to justify
the presence of relatively high-mass nebulae reaching the $M^*$
luminosity, as the empirical invariance of the PNLF bright
cut-off magnitude may imply.

A pragmatic approach to the problem has recently been pursued by
\citet{cia05}, who argued on the possible presence of a blue-straggler
(BS) stellar population in ellipticals, via coalescence of close
binary systems. BSs are commonly observed in Galactic open clusters of
all ages \citep[e.g.][]{kinman,eggen} and in some globulars, too
\citep[e.g.][]{buonanno}, and these objects may in principle reach up
to twice the TO mass, that is $\sim 2~M_\odot$, even in old stellar
systems. While ensuring a convenient fraction of $M^*$ nebulae and
account for the observed $M^*$ invariance, this scenario still leaves,
however, a few open question, which we address below.

\begin{figure}
\centerline{
\psfig{file=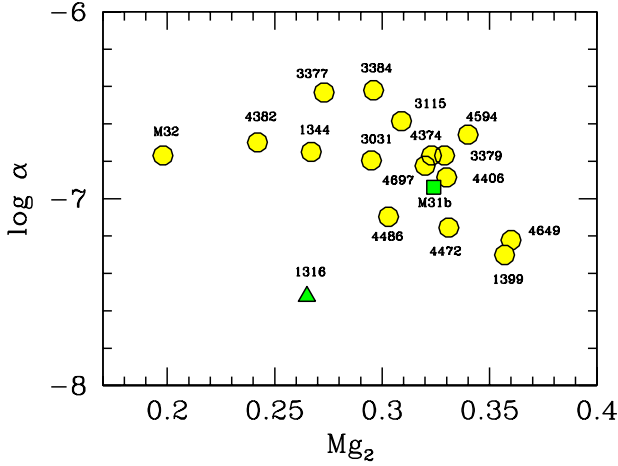,width=\hsize,clip=}
}
\caption{The observed distribution of the elliptical galaxy sample of
Table~\ref{e-sample} (plus M~32 and the bulge of M~31, as labeled on
the plot) versus Lick spectrophotometric index Mg$_2$.  Note the
relative lack of PNe (per unit galaxy luminosity) in more metal rich
ellipticals.  The relevant case of the merger galaxy NGC~1316 is
singled out, while the two active star forming ellipticals NGC~205 and NGC~5102 are
out of range with Mg$_2 \lesssim 0.1$ and not shown. See text for a discussion.}
\label{mg2}
\end{figure}

\subsection{Diagnostic planes}
\label{diagnostic}

\begin{figure}
\centerline{
\psfig{file=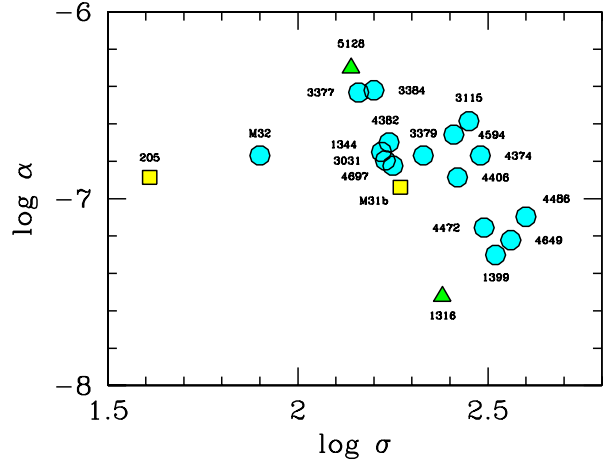,width=\hsize,clip=}
}
\caption{Same as Fig.~\ref{mg2}, but for the galaxy velocity
dispersion $\sigma$ in km~s$^{-1}$.  It is evident a lower value of
$\alpha$ in high-$\sigma$ (roughly more massive) galaxies. See text
for further details.}
\label{sigma}
\end{figure}

\begin{figure}
\centerline{
\psfig{file=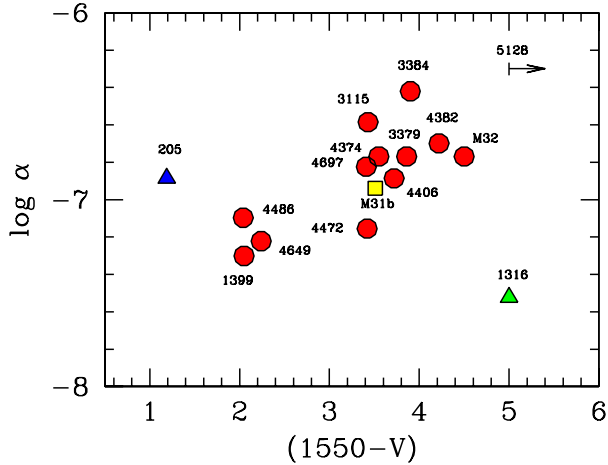,width=\hsize,clip=}
}
\caption{The luminosity-specific PN number versus ultraviolet color
$(1550-V)$, as originally defined by \citet{burstein}, for the
elliptical galaxy sample of Table~\ref{e-sample} (plus the Andromeda
satellites and the bulge of M~31).  Some relevant cases, like NGC~205
and NGC~5102 (star forming), NGC~1316 and NGC 5128 (merger ellipticals) are singled
out on the plot. Note the tight relationship between ``quiescent''
ellipticals and $\alpha$, with UV-bright galaxies to be also
PN-poor. See text for a full discussion of this important effect.}
\label{uvcol}
\end{figure}

To evaluate the impact of the different AGB and PAGB evolutionary pictures in elliptical 
galaxies, it may be useful to investigate the correlations of 
$\alpha$ with other observed quantities for the galaxy sample of Table~\ref{e-sample}, 
as displayed in Fig.~\ref{mg2}, \ref{sigma} and \ref{uvcol}.  

The plot of the Lick Mg$_2$ index \citep{faber} in
Fig.~\ref{mg2} is of special interest, in this regard, because it is the least affected by
reddening. Here, the $\alpha$ correlation with galaxy color is more cleanly
replicated, suggesting that metallicity, rather than age, is the relevant
parameter that affects the observed PN rate per unit galaxy luminosity. With the
exception of NGC~1316 (Fornax A), that we shall discuss in some
details, the plot shows that a lower number of PNe per unit galaxy luminosity is produced 
in metal-rich ellipticals.

As chemical and dynamical properties are tightly correlated in
early-type galaxies \citep[][]{fj,terlevich,bfgk}, one may expect some
correlation of $\alpha$ with the galaxy internal velocity dispersion too
(that is, roughly, with the galaxy total mass).  This is shown in
Fig.~\ref{sigma}. Among others, the good fit of the M~31 bulge to the overall correlation
for early-type systems in Fig.~\ref{mg2} and \ref{sigma} could be
considered as an additional piece of evidence of the similarity
between the stellar populations of spiral bulges and ellipticals
\citep{jablonka}.

The general picture of PN evolution, sketched in Sec.~\ref{mcore}, especially 
reflects in the distinctive properties of the early-type galaxy population 
in the ultraviolet spectral range. In particular, the two relevant Post-HB evolutionary paths, 
that lead stars either to a full AGB completion or straight to the high-temperature 
white-dwarf cooling sequence have different 
impact on the galaxy spectral energy distribution (SED) at short wavelength 
compared to the optical luminosity. To explore this important feature,
in Fig.~\ref{uvcol} we plot the $\log \alpha$ values
vs.\ the UV color $(1550-V) = -2.5\,\log [f(1550\,{\rm \AA})/f(V)]$,
as first defined by \citet{burstein}, where the galaxy SED is measured
at 1550~\AA\ and in the Johnson $V$ band. The $(1550-V)$ color, from IUE
observations and corrected for Galaxy extinction, was provided by \citet{burstein} for
a fraction of galaxies in Table~\ref{e-sample}.\footnote{The $(1550-V)$ color for 
NGC~1316 and NGC~3384 has also been added to Table~\ref{e-sample}, as obtained from 
the \citet{rifatto} UV catalog. These entries, however, are reported with a larger error, 
as the galaxy flux at 1550~\AA\ derives from a crude extrapolation of the
reported magnitudes at 1650~\AA\ and 2500~\AA.}

Figure~\ref{uvcol} shows a tight correlation between the value of
$\alpha$ and the ultraviolet emission, with 
massive UV-enhanced ellipticals, like NGC~4649 and NGC~4486 in Virgo, or NGC~1399 
in Fornax that are also much poorer in PNe. To a finer analysis of the plot, one can even 
notice an apparent gap (marginally evident also in Fig.~\ref{sigma}) between these 
three giant ellipticals and the bulk of more ``normal'' galaxies (including the bulge
of M~31).

One may speculate that the bulk of the PN population in ``normal''
ellipticals evolves according to the
case {\it (ii)} scenario; a substantial fraction of stars, in these galaxies, 
would therefore complete its AGB evolution leading to low-mass PNe, with $M_{\rm
core} \leq 0.55$~M$_\odot$. The intervening increase of the
AGB~$\Rightarrow$~Hot-PAGB transition time in this mass range, would eventually shorten
the nebula lifetime and reduce $\tau_{\rm PN}$ from $\sim 30\,000$~yr to $\sim 10\,000$~yr 
or less, as for the stellar population in the M~31 bulge (see Fig.~\ref{templates}).
If the mean PAGB core mass is further decreased in more massive
galaxies, perhaps as a consequence of a stronger mass loss, then one may
expect an increasing fraction of HB stars to feed the {\it AGB-manqu\'e} evolutionary 
channel and fail to produce PNe, as in case {\it (iii)} scenario. 
This would imply an even lower value of $\alpha$
{\it and} a strongly enhanced galaxy ultraviolet emission, as observed
in NGC~4649, NGC~4486 and NGC~1399 in Fig.~\ref{uvcol}.

\begin{figure}
\centerline{
\psfig{file=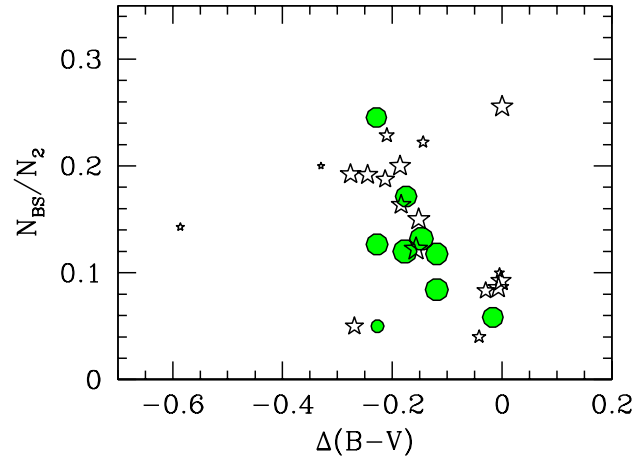,width=\hsize,clip=}
}
\caption{The blue shift of the integrated $B-V$ color of old open clusters in the Galaxy
caused by the BS stellar population, from \citet{xin}. The BS component ($N_{\rm BS}$) 
is normalised in terms of its ratio to the number of MS stars down to 2~mag below the 
TO luminosity ($N_2$). Solid dots are for the oldest ($t \ge 5$~Gyr) clusters, while star markers
include clusters with $1 \le t < 5$~Gyr. Symbol size is proportional to 
cluster statistical richness.}
\label{bs}
\end{figure}

In this framework, the envisaged role of the BS stellar component
could be assessed on the basis of the \citet{xin} analysis of a sample
of old ($t \ge 5$~Gyr) Galactic open clusters. A sizable presence of
BSs is found to severely affect cluster colors, by shifting the
integrated $B-V$ over $\sim 0.2$~mag to the blue (see
Fig.~\ref{bs}). When applied to ellipticals, this argument could place
a quite tight constraint to the overall BS luminosity
contribution, and the size of the induced PN progeny, in order to
avoid unrealistically bluer galaxy colors.  For example, if we take
the case of cluster M~67 (the Praesepe) as a reference, BSs are found
to supply $\sim 50$\% of the total $B$ luminosity \citep{deng},
causing a $\Delta (B-V) = -0.15$~mag shift to the integrated cluster
color \citep{xin}. Converting these values to bolometric luminosities
(e.g.\ by relying on our discussion in Sec.~\ref{bolcol}), this
implies that BS contribution must be {\it well below} $\sim
10$\% of the total galaxy luminosity if we wish to avoid any
measurable effect on the integrated optical colors of ellipticals.

In addition, if $M^*$ PNe in ellipticals really stem from BS
evolution, then one consequence is that the value of $\alpha$ for
these galaxies, as extrapolated from the observation of the very
brightest nebulae alone, could sensibly overestimate the population of
``standard'' (i.e.\ single-star) PNe of lower core mass. According to
eq.~(\ref{eq:scale}), this would imply an even shorter mean lifetime
for single-star nebulae, thus enforcing the importance of the case
{\it (ii)} and {\it (iii)} evolutionary channels.  If two PN
sub-populations of bright (coalesced) PNe and fainter nebulae
generated from standard evolution of low-mass stars really coexist in
early-type galaxies, then one might wonder whether they also
display any distinctive dynamical signature, as the recent case of
NGC~4697 \citep{sambhus} seem to suggest.

To complete the discussion of Fig.~\ref{uvcol}, we now focus on
two major outliers in the plot, namely NGC~205 and NGC~1316. The first case can
easily be assessed as this M~31 dwarf satellite is known to undergo
active star formation \citep[e.g.][]{bica,lee} and its strong
ultraviolet emission (as well as its exceedingly low Mg$_2$ index,
see Table~\ref{e-sample}) is in fact the result of young (a few
$10^8$~yr) MS stars. More special attention, instead, must be paid
to the case of NGC~1316 (Fornax A).

\begin{figure*}
\centerline{
\psfig{file=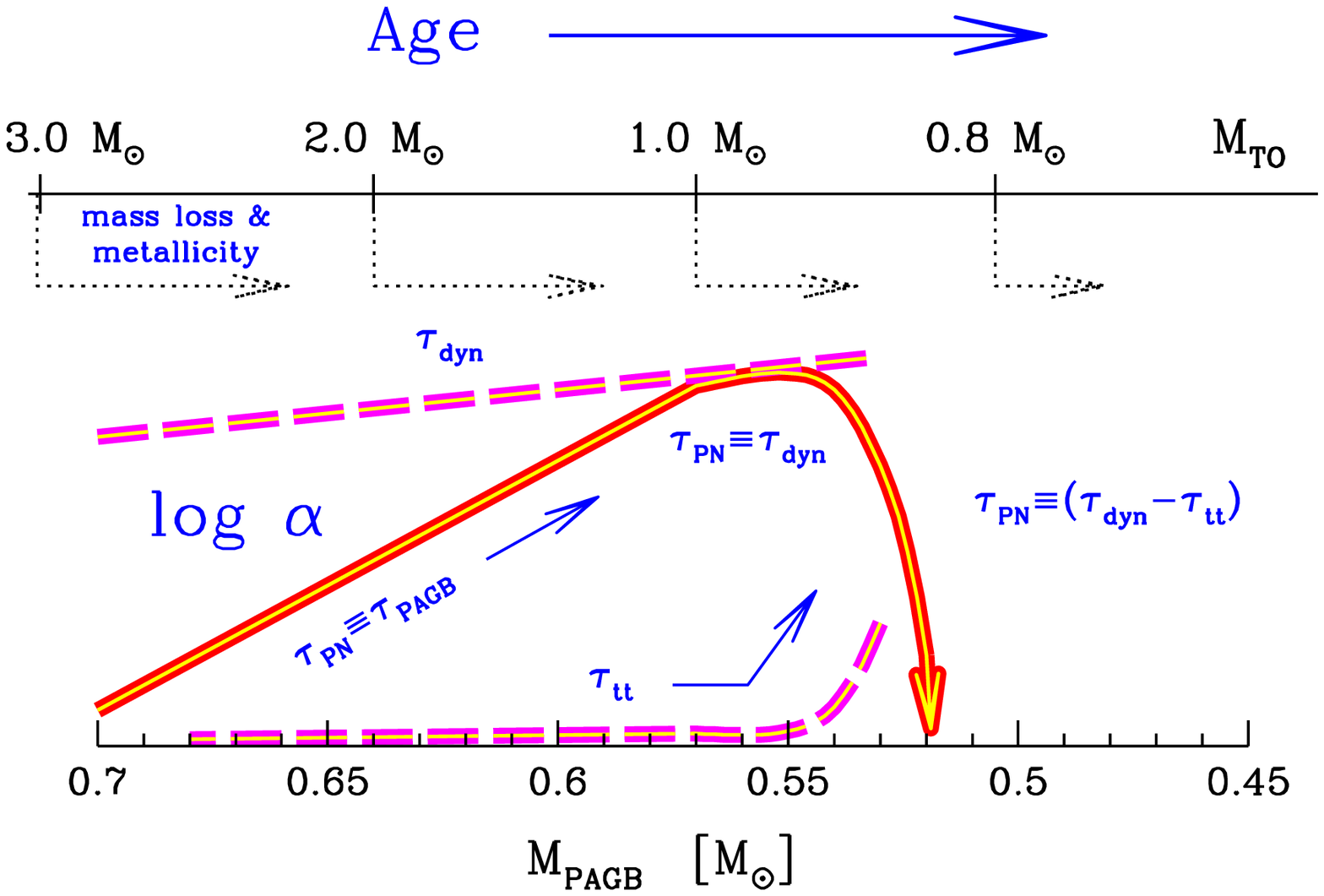,width=0.62\hsize,clip=}
}
\caption{A representation of the envisaged PN evolution versus core
mass of PAGB stars. The effect of different parameters, like
metallicity, mass loss and age is outlined. In particular three
evolutionary regimes are singled out, with PN visibility lifetime
$\tau_{\rm PN}$ (and correspondingly $\alpha$) constrained respectively
by the nuclear timescale ($\tau_{\rm PAGB}$), shell dynamics
($\tau_{\rm dyn}$), and transition time ($\tau_{\rm tt}$). PN
visibility drastically reduces for $M_{\rm core} \lesssim
0.55~M_\odot$ until reaching a critical limit for PN formation about
$M_{\rm core} \simeq 0.52~M_\odot$. See text for full discussion.  }
\label{sketch}
\end{figure*}

\subsection{Galaxy environment and PN evolution: the case of NGC~1316 and Cen~A}

Together with NGC~5128 (Cen A),  NGC~1316 is known as one of
the best established examples of ``mergers'' among early-type systems
\citep{schweizer,goudfrooij}. The study of these two galaxies may
therefore lead to a preliminary assessment of the influence of
``active'' galaxy environments on the PN population.

The location of both objects in the $\log \sigma$ plot of
Fig.~\ref{sigma} places them at the extreme {\it (both highest and lowest)}
values of $\alpha$; furthermore, one must also report the peculiar
location of NGC~1316 in the Mg$_2$ plot (Fig.~\ref{mg2}), which shows a
severe PN deficiency when comparing for instance with NGC~4382, of
similar mass and metallicity.\footnote{Unfortunately, the lack of
Mg$_2$ data for NGC~5128 prevents a similar comparison for this
galaxy.}  Regarding the ultraviolet properties of these ``merger
templates'', the upper limit to the 1540~\AA\ flux for NGC~5128, as
reported by the NED database, suggests a presumably very ``red'' UV
color for this galaxy, such as $(1550-V)_o \gg 5$, in line with the
observations of NGC~1316.

Definitely, in spite of similar observed spectrophotometric properties (i.e\ $\log
\sigma$, $(B-V)_o$ and, likely, the $(1550-V)_o$ color,
cf. Table~\ref{e-sample}), it remains difficult to understand the role 
of galaxy merging mechanisms that led, for Cen~A and For~A, to such a dramatically 
different behaviour in terms of PN population.

\section{Discussion and conclusions}\label{end}

Wide-field CCD detectors, and accurate
selection criteria based on H$\alpha$ and [O{\sc iii}] narrow-band
photometry, allow a systematic survey of the PN population
both in the Milky Way and in external galaxies. Deep observations now
explore the PN luminosity function several magnitudes below the bright
cut-off, as recently achieved in the SMC \citep{jacoby05}.  The
corresponding developments in the theoretical modeling help in
clarifying the main physical mechanisms that constrain the nebular
properties, especially those related to shell dynamics
\citep[e.g.][]{villaver,perinotto04} and chemical composition
\citep{liu, perinotto}, although a clear understanding of the
the PNLF at bright and low magnitudes is still to come.

One open question deals with the empirical evidence for a nearly
constant absolute magnitude of the PNLF bright cut-off ($M^*$), which
makes bright PNe effective standard candles for the intermediate
cosmic distance scale \citep[$\lesssim 100$~Mpc;][]{jacoby89,ciardullo}.  
This feature is not explained by the theoretical models, which predict $M^*$ to be a
function of age, metallicity and other distinctive properties of the
parent stellar population \citep{marigo04}.  In particular, this
problem is stronger for elliptical galaxies, where a lower TO mass for
their stellar populations may hardly reconcile with
the required presence of $\sim 2~M_\odot$ stellar progenitors for the $M^*$
nebulae \citep{cia05,marigo04}.

The PNLF faint-end tail is also subject to a substantial uncertainty,
with observations showing a decrease in the number of PN at about
6~mag fainter than $M^*$ \citep{jdm}, and theory which predicts a
distribution reaching magnitudes as faint as 8~mag below the PNLF
bright cut-off.  We have discussed in Sec.~\ref{lgcensus} that our
limited knowledge of the PNLF faint end reflects in a factor of two
uncertainty on the estimated total number of nebulae for a given
stellar system.  The errors become larger for the most distant
galaxies, external to the LG. In most cases, the PNLF is only sampled
in its brightest magnitude bin, and the observed PNe number must be
multiplied by a factor of 20-40 (see Table~\ref{lf}) to provide an
estimate for the whole PN population.

Despite these large uncertainties, the study of the
luminosity-specific PN number (the so-called ``$\alpha$'' parameter in
our discussion) in external galaxies has allowed us to tackle the
properties of the PN population in different environments, from dense
bulge-dominated galaxies to very low density stellar populations such
as those of the intracluster diffuse stellar component
\citep{aguerri,feld04}.  From the theory presented in
Sec.~\ref{sec:s2} and \ref{ell}, there is a close {\it liason} between the
value of $\alpha$ and the PN visibility lifetime: as the specific
evolutionary flux, $\cal B$, in eq.~(\ref{eq:3}) is nearly constant, then
\begin{equation}
\tau_{\rm PN} \approx 30\,000~(\alpha/5.4\,10^{-7})~{\rm yr}.
\label{eq:stima}
\end{equation}
Depending on the stellar core mass at the beginning of the AGB phase,
the value of $\tau_{PN}$ must be compared with three relevant
timescales that determine the PN evolution. As sketched in Fig.~\ref{sketch},
they are the PAGB core lifetime ($\tau_{\rm PAGB}$), the dynamical
timescale for the nebula evaporation ($\tau_{\rm dyn}$), and the
transition time for the stellar core to leave the AGB and reach the
high-temperature regime required to trigger the nebula [OIII] emission
($\tau_{\rm tt}$).  Along the SSP evolution, mass loss eventually settles
the absolute clock that links the $\tau_{\rm PN}$ evolution with the
SSP age (see Sec.~\ref{mc}).

As far as core evolution provides the leading timescale for the PN
visibility, standard mass loss theory {\it \`a la} \citet{reimers}
predicts that $\tau_{\rm PN}$ (and correspondingly $\alpha$) {\it
should increase with SSP age}.\footnote{For a characteristic value of
the mass loss parameter $\eta$, we have from \citeauthor*{iben} that
the final mass of PN nuclei roughly scales as $M_{\rm PAGB} \propto
\eta^{-0.35}\,t^{-0.29}$, recalling that, to a first approximation, $t
\propto M_{\rm TO}^{-3.5}$ \citep{b02}.  According to
footnote~\ref{nota1}, this eventually leads to $\tau_{\rm PAGB}
\propto \eta^{2.2}\,t^{1.8}$.}  More generally, one would expect this
to be the case of late-type galaxies, where star formation is on-going
and a consistent fraction of high-mass stars is younger than a few Gyr
(cf.\ Fig.~\ref{alpha_ssp}). As the fraction of young stars is higher
in irregulars than in elliptical galaxies, and the integrated $B-V$
color becomes bluer along the E~$\to$~Im Hubble morphological
sequence, then {\it one may conclude that lower values of $\alpha$ are
expected in bluer galaxies.}

However, observationally this does not occur for the LG galaxy
population, as shown in Fig.~\ref{templates}. In fact, while the
suggested theoretical range for $\alpha$ in the $\eta = 0.3$ models
matches the observations {\it on average}, the latter points
to a constant behaviour with galaxy morphological type, with a typical
rate between 1 and 6~PNe per $10^7$~L$_{\odot}$.  From these
observations, the inferred PN lifetime in LG spirals and irregulars
should exceed 10\,000~yr, a value that requires the presence of a
substantial fraction of stars with $M_{\rm core} \lesssim 0.65~M_\odot$
(cf.\ Fig.~\ref{mcr}) even in those galaxies with the most active star
formation activity.

The relatively low final mass and the tight mass range required by the
``dying'' stars in external galaxies find independent confirmation in
the Milky Way, as also indicated by the observational evidence of the
IFMR for local open clusters by \citeauthor*{weidemann}.  We have
shown in Sec.~\ref{lgcensus} that population synthesis models using
this empirical IFMR produce a better fit (both in absolute value and
relative trend with galaxy morphological type) for the correct value of PN 
density per unit galaxy
luminosity in the LG galaxies, along the whole Hubble morphological
sequence (see Fig.~\ref{templates}).  This result indicates that the
dynamical evolution plays a central role in setting the overall PN
observed properties as for low core-mass stars, the dynamical rather
than nuclear timescale is the real driving parameter to constrain PN
visibility (see Fig.~\ref{sketch}). As a consequence, one may state that,
{\it rather than probing any real mass distribution of PAGB stars, the
PNLF basically tracks the time evolution of the expanding shell around stellar
nuclei of nearly fixed mass} \citep{henize}.

Within this picture, one problem is related with the PN evolution in
(giant) elliptical galaxies. Here, the empirical evidence for a constant
PNLF bright cut-off magnitude would require a sizable component of
relatively high-mass stars, well above the TO mass of $\sim 1~M_\odot$
expected for these old stellar systems. Following \citet{cia05}, BS stars
originating from coalesced close binary systems according to a classical
evolutionary scheme \citep{mccrea}, may possibly overcome the dilemma.
However, if ellipticals do indeed host a BS population in a fraction similar to
what we detect in old Galactic open clusters, then galaxies 
should appear roughly 0.2~mag bluer than observed, in $B-V$ (see Fig.~\ref{bs}). 
As a consequence, from our arguments in Sec.~\ref{diagnostic}, one is led to conclude 
that, in any case, BSs cannot provide much more than a few percent of galaxy 
bolometric luminosity and {\it PNe coming from the BS progeny must be confined to the
very brightest bin of the PNLF, thus representing a marginal fraction of
the global PN population.}

The relative lack of massive stars in early-type galaxies, and a high (super-solar?)
metallicity possibly easing a stronger mass loss in these galaxies, might actually 
lead to a larger component of blue HB stars, eventually evolving
into low-mass PNe.  In Sec.~\ref{mcore}, we have seen that for the
low-mass cores ($M_{\rm core} \lesssim 0.55~M_\odot$), the increase in
the transition time $\tau_{\rm tt}$ reduces effectively the visibility
timescale of the nebula (see, again Fig.~\ref{sketch}), thus
decreasing the value of $\alpha$. This trend is expected to depend on
galaxy color (or metallicity, as shown by the Mg$_2$ distribution of
Fig.~\ref{mg2}) with a lower PN density per unit galaxy luminosity for
the redder ellipticals.

Furthermore, for the stellar component with $M_{\rm core} \lesssim
0.52~M_\odot$, both HB and Post-HB evolution occur at high temperature
($T_{\rm eff} \gg 10^4$~K) thus skipping the PN phase entirely
\citep{ct,dorman,blocker}.  For these stars, the {\it AGB manqu\'e}
evolution can effectively transfer some fraction of the PAGB energy
budget to the hot HB tail and affect the integrated galaxy SED. As a
consequence, one may expect that the most UV-enhanced ellipticals
display also the lowest values of $\alpha$.  We have shown in
Sec.~\ref{ell} that such a tight correlation is observed for
elliptical galaxies in the Virgo cluster and other groups (see
Fig.~\ref{uvcol}). Recent HST observations of the resolved stellar
population of M~32 and the bulge of M~31 \citep{brown00,brown98}
further support this proposed scenario.

The presence in our sample of two merger galaxies (Fornax~A and Cen~A)
may provide the opportunity to study the effect of galaxy interactions
on the PN population. Based on the available data however, no firm
conclusions can be drawn, although one has to remark that these two
galaxies stand out as those with the most extreme (both highest and
lowest) values of $\alpha$ in our sample.  Perhaps such a huge
variation in the galaxy PN population may be related to the disruptive
effect of the intergalactic ram pressure, as recently considered by
\citet{villaver2} and \citet{villa05}.

\section*{Acknowledgments}
We would like to thank Robin Ciardullo, the referee of this paper, for 
enlightening comments on the role of blue stragglers in PN evolution of
elliptical galaxies. Giuseppe Bono, Ortwin Gerhard, Laura Greggio,
Detlef Sch\"onberner and Letizia Stanghellini are also acknowledged for 
useful discussions and suggestions on earlier drafts of this work.  
This project received financial
support by INAF, under grants PRIN/02 (PI: MA) and PRIN/05 (PI: AB), 
and the Swiss National Foundation. Our analysis has made extensive use of
different on-line extragalactic databases, namely the NASA/IPAC
Extragalactic Database (NED), operated by JPL/CIT under contract with
NASA, the Hyper-Linked Extragalactic Databases and Archives
(HyperLeda) based at the Lyon University, and the VizieR catalog
service of the Centre de Donn\'ees astronomiques de Strasbourg.

\label{lastpage}

\bsp
\end{document}